\documentclass[12pt,preprint]{aastex}

\slugcomment{Submitted to the Astrophysical Journal - Feb. 19, 2004.}

\shorttitle{An Infrared View of T Tauri}
\shortauthors{Beck et al. }

\begin{document}


\title{A High Spatial Resolution Infrared View of the T Tauri Multiple System}


\author{Tracy L. Beck\altaffilmark{1,2,3} G. H. Schaefer,\altaffilmark{2} M. Simon,\altaffilmark{2,3} L. Prato,\altaffilmark{3,4}, J. A. Stoesz,\altaffilmark{5} and R. R. Howell\altaffilmark{6}}

\email{tbeck@gemini.edu}


\altaffiltext{1}{Gemini Observatory, 670 N. A'ohoku Pl. Hilo, HI 96720}
\altaffiltext{2}{Department of Physics \& Astronomy, SUNY-Stony Brook, Stony Brook, NY 11794-3800}
\altaffiltext{3}{Visiting Astronomer at the
                              Infrared Telescope Facility,
                              which is operated by the
                              University of Hawaii under
                              Cooperative Agreement no. NCC
                              5-538 with the National
                              Aeronautics and Space
                              Administration, Office of Space
                              Science, Planetary Astronomy
                              Program}
\altaffiltext{4}{Department of Physics \& Astronomy, UCLA, Los Angeles, CA 90095}
\altaffiltext{5}{Department of Physics and Astronomy, University of Victoria, Victoria, BC V8W3P6, Canada}
\altaffiltext{6}{Department of Geology \& Geophysics, University of Wyoming, P.O. Box 3905, University Station, Laramie, WY 82071}


\begin{abstract}

We present the results of our monitoring study of the IR photometric and spectroscopic variability of the T Tau multiple system.  We also present data on the apparent position of T Tau S with respect to T Tau N, and two new spatially resolved observations of the T Tau Sa-Sb binary.  T Tau N  has not varied by more than 0.2 magnitudes in K and L$'$ flux during the 8 years of our observations, though its Br$\gamma$ and Br$\alpha$ emission line fluxes have varied by nearly a factor of four during this time.  For the unresolved T Tau S system, we have derived a 20 year light curve based on our images and on measurements available in the literature.  T Tau S varies by 2-3 magnitudes in K and L$'$-band brightness in a ``redder when faint'' manner, consistent with changes along the line of sight in the amount of material that follows an ISM extinction law.  Absorption in the 3.05$\mu$m water-ice feature is seen only in the spectra of T Tau S and it displays variations in depth and profile.  H$_2$ (2.12$\mu$m) emission is also detected only at the position of T Tau S; the H$_2$, Br$\gamma$ and Br$\alpha$ emission line fluxes also vary.   We have found that the apparent positions of T Tau S with respect to T Tau N and T Tau Sb with respect to Sa are consistent with gravitationally bound orbital motion.  The possible orbits of the T Tau S binary imply that Sa is likely the most massive component in this young triple.  A reanalysis of the motion of the radio source associated with T Tau S provides no evidence for an ejection event in the T Tau system.

\end{abstract}




\section{Introduction}

T Tauri is one of the most vigorously studied stars in the northern sky, but the nature of this young multiple system remains a mystery.  Joy (1945) used T Tau as the prototype of a new class of stars that show significant variability, are associated with bright or dark nebulosities and have characteristic emission line spectra.  Ambartsumian (1947; 1949) first proposed that the T Tauris are in fact young solar type stars in an early stage of formation.  T Tau is a $\sim$1 Myr old K0 star with a mass of $\sim$2 M$_{\odot}$ and a visual extinction of $\sim$1.5 magnitudes (Cohen \& Kuhi 1979; White \& Ghez 2001).

Dyck, Simon \& Zuckerman (1982) discovered that T Tau has an infrared companion $\sim$100 AU to the south.  This source, T Tau South (T Tau S), has never been detected at optical wavelengths to a limiting magnitude of $\sim$19.6 at V (Stapelfeldt et al. 1998), yet it dominates the system emission at wavelengths longward of 3$\mu$m.  T Tau S has an integrated luminosity twice that of T Tau North, the optical star (T Tau N), and shows greater than 2 magnitudes of variability in the infrared (Ghez et al. 1991; Roddier et al 2000; Beck, Prato \& Simon 2001).  K-band spectral observations confirm that T Tau N is a young K star with significant continuum veiling and hydrogen Br$\gamma$ emission indicative of ongoing accretion activity (2.16 $\mu$m; Beck, Prato \& Simon 2001; Kasper et al. 2002;  Prato \& Simon 1997; Muzzerole et al. 1998).  The spectrum of T Tau S rises steeply toward longer wavelengths, shows no photospheric absorption features characteristic to young solar type stars, and shows weak and possibly variable Br $\gamma$ and H$_{2}$ (2.12 $\mu$m) emission (Beck, Prato \& Simon 2001; Kasper et al. 2002; Duch\^ene, Ghez \& McCabe 2002). 

T Tau S is the prototype of a young class of stars known as infrared companions (IRCs) (Zinnecker \& Wilking 1992; Koresko, Herbst \& Leinert 1997; Koresko \& Leinert 2001).  IRCs show significant variability in infrared flux, often have larger luminosities than their optical counterparts, and are rarely or never seen at wavelengths shortward of 1.2 $\mu$m.  Their spectral energy distributions are best fit by blackbodies with temperatures less than 1200 K, which is too cool for a stellar photosphere.  If they were observed as single stars, IRCs would likely be defined as ``Class I'' or ``protostellar'' T Tauris only a few times 10$^{5}$ years old.  However, in most cases this classification is inconsistent with the ages of their optical counterparts.   Koresko, Herbst \& Leinert (1997) proposed that IRCs are relatively normal young stars that are obscured by their circumstellar material and are experiencing variable accretion episodes. 

T Tau S was first shown to vary in flux when Ghez et al. (1991) discovered a gray $\sim$2 magnitude flare in the infrared.  Roddier et al. (2000) detected T Tau S in the J-band when it was $\sim$6 magnitudes fainter than T Tau N; this remains the shortest wavelength in which T Tau S has been observed.  Its near infrared spectral energy distribution (SED) is best fit by a black body of temperature $\sim$800K (Koresko, Herbst \& Leinert 1997), and the pronounced level of polarized light in the K-band supports the idea that T Tau S is a Class I protostar (Kobayashi et al. 1997).  However, Koresko, Herbst \& Leinert (1997) showed that T Tau S is best modeled as a 2.5 M$_{\odot}$ T Tauri star observed through 35 magnitudes of visual extinction from circumstellar material.   The infrared spectra of T Tau S shows prominent absorption features of water-ice at 3.05$\mu$m and silicates at 9.7$\mu$m, which are known to correlate with the visual extinction through molecular clouds.  Applying the relations between the optical depths of these features and the visual extinction along the line of sight suggests that T Tau S is obscured by only 5-10 magnitudes in the visible (Beck, Prato \& Simon 2001; Ghez et al. 1991; Herbst, Robberto \& Beckwith 1997; Whittet et al. 1988; Teixeira \& Emerson 1999).

Using speckle holographic observations, Koresko (2000) discovered that T Tau S is itself a binary with a companion at a projected separation of $\sim$0.$''$05 ($\sim$ 7 AU).  Spatially resolved spectra reveal that this companion (T Tau Sb) is a young star of spectral type M1 observed through $\sim$8 magnitudes of visual extinction and that the IRC (T Tau Sa) is featureless except for Br $\gamma$ emission (Duch\^ene, Ghez \& McCabe 2002).   Reipurth (2000) proposed that the infrared companion phenomenon is a direct result of the dynamical interaction of young stars in a triple system.  Based on this hypothesis, all IRCs, or their optical counterparts, should have a companion.  In addition to T Tau, several IRC systems have been identified as triples or higher order multiples (Duch\^ene et al. 2003; Koresko 2002; Ressler \& Barsony 2001; Brandeker et al. 2001). 

Both T Tau N and S are radio sources, but the location and movement of the radio source associated with T Tau S is not consistent with the infrared positions (Johnston et al. 2003; Roddier et al. 2000; Ghez et al. 1995).  Further analysis of the motion of the radio source coupled with the known motion of T Tau S in the infrared has led Loinard et al. (2003) to suggest that T Tau Sb is being ejected from the system.  However, Duch\^ene, Ghez \& McCabe (2002) and Furlan et al. (2003) have found that the infrared position of T Tau Sb with respect to T Tau Sa is consistent with orbital movement of the components.  The data show interesting and contradictory behavior, and a definitive relationship between the radio emission and the positions of T Tau Sa and Sb has not yet been established.

In the hope that the significant variations in infrared flux and the apparent motions of the system components can provide clues to the nature of T Tau S and the IRC phenomenon as a whole, we have monitored the photometric and spectroscopic variability of the T Tau system.  We provide the details of our observations in \S2 and present the observational results of the variability in \S3 and orbital motion in \S4.  In \S5 and \S6 we discuss the nature of the components in this enigmatic young triple.  A summary appears in \S7.

\section{Observations}

Many of the imaging and spectroscopic observations for this project were obtained through scheduled and service observations at NASA's Infrared Telescope Facility (IRTF) on Mauna Kea.  These observations resolved T Tau N and S, but not T Tau Sa and b.  For the presentation and analysis of our results (Tables 1, 3, and Figures 1-8) we refer to the combined flux of the T Tau S binary as T Tau S.  Unresolved observations of T Tau N and S were also made at the Wyoming Infrared Observatory (WIRO).   Two high spatial resolution adaptive optics observations of the system have been made at the W. M. Keck and Gemini North Observatories.  Table 1 presents a log of the observations obtained for this study.  

\subsection{Imaging Observations}

Our spatially resolved observations of the T Tau N-S system were carried out from August 1995 to January 2002.  At the IRTF, we used the infrared camera NSFCam, which is equipped with a 256 $\times$ 256 InSb array, with the 0.$''$056/pixel plate scale.  The K and L$'$ observations of the T Tau system were made by taking 100 0.1 second ``speckle-mode'' exposures to freeze the seeing.  Subsequent shift-and-add analysis produced nearly diffraction limited images of the 0.$''$7 T Tau system.  The fluxes were determined from the total system flux and the flux ratio in the derived images.  In September to early October 1999 we obtained unresolved photometry of T Tau  at WIRO using IoCam1, which has a 62 $\times$ 58 InSb array with a plate scale of 0.$''$33/pixel.   For these observations, the flux of T Tau S was derived from the total flux by assuming the flux of T Tau N was constant (see \S 3).   

All observations obtained at the IRTF were reduced by determining the point spread function (PSF) at each of the observed wavelengths by making shift-and-add images from a series of short exposures of the photometric standards.  The flux ratios of T Tau N and S at the observed wavelengths were derived with both $\chi^{2}$ and cross-correlation binary fitting procedures using this PSF; the results were always consistent to within 1$\sigma$.  

Our adaptive optics observations from the W.M. Keck and Gemini North Observatories resolved the $\sim$0.$''$1 separation T Tau S binary.  In both cases, T Tau N was used for adaptive optics wavefront correction.  The data from 2002 Oct 30 were obtained at the W. M. Keck Observatory with NIRC2, which uses a 1024$\times$1024 Aladdin-3 InSb array (Wizinowich et al. 2000).  We used the narrow-field camera with a plate scale of 0.$''$0099 $\pm$0.$''$0005/pixel and orientation PA=0.7$\pm$0.2, as measured during commissioning of NIRC2 (R. Campbell, private comm.).  The position angles and their uncertainties reported in section 4 include this offset and uncertainty added in quadrature.  The data were taken as 15 images in a 5 point dither pattern in each of the H, K$'$ and Br$\gamma$ filters.  Each individual image consisted of 10 coadds of 0.18 seconds in the H and K$'$-bands with 154 Hz AO correction, and 0.3 seconds in the Br$\gamma$ filter using 358 Hz AO correction.  Conditions were not photometric at the time of the observations, and the improved guide correction for the Br$\gamma$ observations resulted from the clearing of cirrus clouds.  For the subsequent data reduction, the pixel scale was expanded by a factor of three through a cubic-convolution interpolation to improve the precision of the stellar positions.  The close pair, Sa-Sb, was modeled over a grid of brightness ratios and separations using T Tau N as the PSF.  The best-fit solution was determined by the minimum variance between the model and the observations.  Details of this reduction method are described in Schaefer et al. (2003).

We also obtained high spatial resolution images of T Tau on 2002 Dec. 20 during comissioning observations of the Altair adaptive optics system at Gemini North Observatory (Herriot et al. 2000; V\'eran et al 2003).  We used the near infrared imager (NIRI; Hodapp et al. 2003) with the f/32 configuration which has a plate scale of 0.$''$0218/pixel and an orientation of PA=0.0$\pm$0.05; this was measured during commissioning of NIRI+Altair.  These observations were taken in photometric conditions, with natural seeing of $\sim$0.$''$4 to 0.$''$5 at 500 nm.  Adaptive optics corrections were applied at a rate of 1 kHz for H, K and narrow-band H$_{2}$(1-0) filter observations.  To acquire the data we used a 9-point dither pattern with 0.$''$2 offsets, and the final coadded images summed to 81 seconds in the K-band, 61 seconds at H and 108 seconds in molecular hydrogen.  The images were shifted and stacked using the {\c stsdas.drizzle} package in IRAF\footnote{IRAF is distributed by NOAO, which is operated by AURA, Inc., under cooperative agreement with the NSF. }.  The scale was shrunk by 70\%, and the data were drizzled into an array that oversampled the original pixels by a factor of 2.  T Tau N was saturated in these images, so the separation and flux ratio of the T Tau S binary was determined using tools in the IRAF package.   The Altair K-band image of T Tau is presented and discussed in section 3.3.

\subsection{Spectroscopic Observations}

We obtained five spectra of T Tau using SpeX, the facility near infrared spectrometer at the IRTF (Rayner et al. 2003).  The SpeX instrument consists of a 1024$\times$1024 InSb array, Bigdog, which has a plate scale of 0.$''$15/pixel and is used for spectroscopic observations, and a 512$\times$512 InSb array, Guidedog, which has a 0.$''$12/pixel plate scale and serves as a guiding/imaging camera.  The spectra were obtained using Bigdog with a 0.$''$5 slit and a 15$''$ beam-switched mode for sky correction.  We used the long wavelength cross-dispersed setting, LXD1.9, which provides simultaneous spectral coverage of the 1.9-4.1 $\mu$m region.  The resulting spectra have resolution R $\sim$ 1100.  In addition to T Tau, we also observed a nearby A0 spectral type star to correct for telluric absorption and instrumental effects and to serve as a PSF calibrator to extract the blended spectra of T Tau N and S.  We removed the photosperic hydrogen features present in the A0 spectral calibrator using the ``SpeXTool'' data reduction software with the deconvolution method described by Vacca et al. (2003).  In all cases, we were able to obtain good correction in the Hydrogen photospheric features using this method.

The spectral images were flat fielded and beam-switch sky subtracted.  For each of the 5 orders in the spectroscopic data frames, a 1024$\times$25 pixel subarray was extracted.  One dimensional PSFs, 25 pixels long, were constructed by median filtering 11 columns in the cross dispersion direction of the calibration star subarrays (see also Beck, Prato \& Simon 2001 and Prato, Greene \& Simon 2003).  The spectra of the binary components were determined by fitting model binaries, created from overlapping PSFs, to the observed data in each of the cross dispersed spectral columns, constrained by the known separation of T Tau N and S.  The individual spectral amplitudes were determined by the model PSF binary which yielded the best fit in the least-squares sense.  The typical residuals in our PSF fitting technique were $\sim$3-9\% per pixel of the peak flux.  After the five orders of spectra were extracted for T Tau N and S, we copied these data into a format suitable for further reduction with the ``SpexTool'' data reduction software available from the IRTF (Cushing et al. 2003).  The SpexTool tasks xtellcor, xmergeorders and xcleanspec were used for telluric correction, to merge the orders into a single spectrum, and to clean bad pixels.   We cleaned the A0 spectral calibrator

To flux calibrate the spectra, images of T Tau and the photometric standard star, 16 Tau, were made in the K and L$'$-bands using the Guidedog imaging array.  A standard circular aperture photometry procedure was used with an aperture radius of 3$''$.  The total flux and the flux ratio derived from the spectroscopic observations were coupled to calibrate the spectra.   Figure 1 shows the final corrected, merged, and cleaned spectra for T Tau N and S from observations obtained on 2001 Oct. 15.  Atomic hydrogen and photospheric absorption features typical of K0 stars are present in the spectrum of T Tau N, and atomic and molecular hydrogen emission is apparent in T Tau S.  T Tau S also shows a broad absorption feature at 3 $\mu$m which is caused by stretch-mode oscillations of water-ice frozen onto cold dust grains.  In all observations of these spectra, the water-ice absorption is only seen in T Tau S, and the photospheric absorption features are only detected in T Tau N.  This distinction suggests that our spectral extractions are successful at separating the flux from the individual components in the T Tau N-S system.




\section{Variability Results}

Photometric monitoring that resolves T Tau N and S has continued since the discovery of T Tau S in 1981.  The results of these studies are summerized in Table 2.  The data that we have accumulated doubles the number of published resolved observations of the T Tau system in the K-band, and increases the number of observations in the L$'$-band by a factor of six.

\subsection{The Infrared Variability of T Tau N}

Figure 2 presents the light curve of T Tau N in the K and L$'$ photometric bands.  The star has not varied significantly ($>$0.2 mag) during the course of these observations and has average magnitudes of K = 5.53$\pm$0.03 and L$'$=4.32$\pm$0.05.  Hence, for our unresolved observations of the T Tau system and for flux ratios of T Tau N and S that are reported in the literature, the K and L$'$ magnitudes of T Tau S are determined using these average values for T Tau N with conservative uncertainties of $\pm$0.1 at K, and $\pm$0.15 at L$'$ (see Table 2). 

In Figure 3 we present K-band (2.0-2.4$\mu$m) spectra of T Tau N; five spectra were obtained with SpeX (Table 1 and \S2.2) and 1 with KSPEC on the UH 2.2 meter telescope in 1996 Nov. 3 (previously published in Beck, Prato \& Simon 2001).  These spectra show prominent Br$\gamma$ emission and photospheric absorption lines of Na (2.208 $\mu$m) and Ca (2.264 $\mu$m).  T Tau was among the first young stars discovered to have H$_{2}$ S(1) 1-0 (2.12$\mu$m) emission (Beckwith et al. 1978).  We do not detect this emission feature in the spectra of T Tau N, so we interpret this to mean that a significant component of the 2.12$\mu$m emission is associated with T Tau S and/or its circumstellar material rather than with T Tau N.  Table 3 presents the Br$\gamma$ and Br$\alpha$ fluxes of T Tau N and 3$\sigma$ detection limits for the H$_{2}$ emission.  Although the K and L$'$-band fluxes of T Tau N do not vary on a statistically significant level, the Br$\gamma$ and Br$\alpha$ emission fluxes have varied by more than a factor of 3.

The spectra observed in 1996 Oct., 2000 Nov. 18 and 2001 Oct. show the CO $\Delta\nu=$2 bands clearly in absorption toward T Tau N; the CO {\it v}=2-0 feature at 2.29$\mu$m typically has a depth greater than that of the Ca absorption, consistent with the $\sim$K0 spectral type of T Tau N.  The other spectra have slightly lower signal-to-noise, thus it is not possible to identify with any confidence the CO absorption in the 2000 Oct., 2000 Nov. and 2002 Jan. spectra of T Tau N .  Variability and decreased depth of CO absorption could support the hypothesis that periodic emission arising from the inner circumstellar disk fills in the photospheric features (Greene \& Meyer 1995; Calvet, Hartmann \& Strom 1997; Prato, Greene \& Simon 2003).  Future spatially resolved spectroscopy of T Tau N may provide the opportunity to study variable CO emission and its correlation with the Br$\gamma$ diagnostic of disk accretion activity.

\subsection{The Infrared Variability of T Tau S}
  
Figure 4 shows the light curve of T Tau S in the K and L$'$ photometric bands derived from more than 20 years of spatially resolved observations.  These data are compiled from our own study as well as many measurements reported in the literature (Table 2).  The total amplitude of the observed variability is nearly three magnitudes in the K-band and more than two magnitudes at L$'$.  In the Fall of 1999, the flux of T Tau S increased in both K and L$'$-bands to the brightest level that it has been observed to date; South/North flux ratios were 0.75$\pm$0.06 at K and 2.2$\pm$0.1 at L$'$.  In September 1998 and November 1999 we obtained observations of the system over three consecutive nights; T Tau S did not show any short-term variability during these observations.  The shortest timescale flux variations that we find in T Tau S are from observations made on 06 and 13 Dec. 1997 when it brightened by 0.9 magnitudes in the K-band over only a seven day period.

The K-L$'$ color of T Tau S is plotted versus the K-band magnitude in Figure 5; it includes 35 of our measurements and 3 simultaneous K and L$'$ observations reported in the literature (Ghez et al. 1991, Simon et al. 1996).  This figure demonstrates that T Tau S appears redder in K-L$'$ when faint and bluer when bright.  This color versus magnitude behavior is expected if changes in the amount of obscuring material along the line of sight to the star are responsible for the observed variability.  The best fit to the observed correlation, in the least squares sense, is overplotted in the figure and has a slope of 0.43$\pm$0.11.  For the reddening of material which conforms to an ISM extinction law, i.e., A$_{K}$=0.112A$_{V}$, A$_{L}$=0.058A$_{V}$ (Rieke \& Lebofsky 1985), the slope would be 0.49; this line is also plotted in Figure 5 for comparison.

The similarity between slopes of the ISM extinction law and the best-fit to the data suggests that obscuration plays a dominant role in the observed variability.  The implied extinction changes are large, at least A$_v\sim$20 magnitudes.  Nevertheless, the correlation between K-L$'$ color and K-band magnitude implies that a single mechanism is reponsible for a large part of the photometric variability, and the observed scatter in the fit suggests that other processes play a role also.  We discuss further the possible mechanisms for the variability in \S 6.1.

Figure 6 shows the six K-band spectral observations of T Tau S derived from the same observations that produced the spectra of T Tau N in Figure 3.  In October 1996, weak Br$\gamma$ and H$_{2}$ {\it v}=1-0 S(1) emission features were detected toward T Tau S, at 3$\sigma$ and 2$\sigma$ significance, respectively (Beck, Prato \& Simon 2001).  The five additional observations also show these emission features at varying levels of confidence and thus confirm that molecular hydrogen {\it v}=1-0 S(1) is detected in emission.  Moreover, the molecular hydrogen emission is always observed toward T Tau S; T Tau N shows no evidence for it.   We detect no emission in any of the other molecular hydrogen transitions observable in the 2.0 - 2.4$\mu$m region described by Herbst et al. (1996).  Table 3 gives the fluxes of the Br$\gamma$, Br$\alpha$ and H$_{2}$ lines toward T Tau S.  These features show variability at 5$\sigma$, 4$\sigma$  and 2$\sigma$ levels of significance, respectively.   

In Figure 7 we plot the Br$\gamma$, Br$\alpha$ and H$_{2}$ emission versus the K or L$'$-band flux.  The Br$\gamma$ emission seems correlated with K-band flux; T Tau S has larger Br$\gamma$ flux when bright, and smaller flux when faint (Figure 7a).  Although the Br$\alpha$ line flux also appears to be related to the L-band flux, it is not possible at this time to determine if a correlation exists between these values because fewer measurements are available (Figure 7b).  The H$_{2}$ emission shows no simple correlation with K-band flux (Figure 7c).  Interestingly, the continuum K-band flux from Oct. 2001 to Jan. 2002 decreased by more than a factor of 2, but the H$_{2}$ line flux was constant to within less than 1$\sigma$.  This suggests that the molecular hydrogen emission arises from a component that is extrinsic to the infrared flux variability and hence unaffected by the significant continuum variation.  No photospheric absorption features characteristic of young solar-type stars are observed in the spectra of T Tau S.

Figure 8 shows the five SpeX spectra of T Tau S over their entire range, 2-4$\mu$m.  All of these spectra show the 3.05$\mu$m water-ice feature, as well as emission in the Br$\gamma$, Br$\alpha$ and H$_2$ S(1) lines.  To derive the ice-band optical depths we fit Planck blackbody curves to the continuum flux at 2-2.4$\mu$m and 3.5-4.0$\mu$m.  The best-fit temperatures range from 830 to 1320 K (these values are listed in column 2 of Table 4), but only the 2002 Jan. 2 spectrum is well fit by a single temperature blackbody.  Columns 3 and 4 of Table 4 list the optical depths derived from the spectra and the FWHM of the absorption in cm$^{-1}$.  For comparison, the optical depths derived from narrow-band observations are included in the first three lines of Table 4 (Beck, Prato \& Simon 2001).  

At the present time, relationships between visual extinction and water-ice optical depth or column density, N$_{ice}$, are well determined only for the lines of sight through molecular clouds and for young stars with visual extinctions less than $\sim$10 magnitudes.  Thus, the applicability of these relations to warmer or more dense circumstellar regions is uncertain.  The correlation found by Whittet et al. (1988) relates $\tau_{ice}$ and A$_{v}$: $\tau_{ice}$=[m(A$_{v}$-A$_{v}$(0))], where m=0.093$\pm$0.001 and A$_{v}$(0)=3.3$\pm$0.1.  The visual extinctions found from our spectra using this relation are included in the last column of Table 4.  The $\tau_{ice}$/A$_{v}$ relation from Chiar (1995) gives very similar results.  Teixeira \& Emerson (1999) found a correlation between water-ice column density and A$_{v}$ from spectra of 19 stars seen through and embedded within the Taurus dark clouds:  N$_{ice}$=[(1.0$\pm$0.1)A$_{v}$+(-2.1$\pm$0.8)$\times10^{17}$ cm$^{-2}$.  The water ice column density is derived by N$_{ice}$=$\tau_{ice}$$\Delta\nu_{1/2}$/A, where A=2.0$\times10^{16}$ cm/molecule is the integrated absorption cross section (d'Hendecourt \& Allamandola 1986) and $\Delta\nu_{1/2}$ is the FWHM of the water-ice feature (in cm$^{-1}$).  Columns 5 and 6 of Table 4 list the water-ice column density and visual extinction derived using the Teixeira \& Emerson (1999) relation.   

The visual extinctions derived using Whittet et al.'s (1988) relation are consistently smaller than those derived using Teixeira \& Emerson's (1999).  The uncertainties in the A$_v$ values are also larger, so perhaps the difference between the two relations is not significant.  The K and L$'$ band fluxes have varied by more that a factor of 2 over the time of the spectral monitoring (Table 2) with the change in color consistent with the $''$redder when faint$''$ relation (Figure 5).  However, we do not find any clear correlation between the K or L$'$-band flux level and changes in the optical depth or column density of the water-ice absorption feature.  The spectral shape of the water ice absorption did become broader and shallower from 2001 Oct. 15 to 2002 Jan. 2.  Similar differences in the water-ice absorption profiles are seen when a range of stars are observed (Smith, Sellgren \& Tokunaga 1988; 1989).  Both theoretical models and laboratory research show that narrow, more sharply peaked absorption profiles are characteristic of warmer crystalline ices on the surfaces of grains, while the broad, more rounded profiles are indicative of amorphous ices with temperatures less than $\sim$110K (Hobbs 1974; Smith, Sellgren \& Tokunaga 1988; 1989; Maldoni et al. 1998).  Our observations of T Tau S present very tentative evidence of variability in profile shape, speculatively suggesting the processing of the water-ice on the surface of cool dust in the circumstellar environment.  However, we want to emphasize that the variability in the profile shape is significant only at a $\sim$3$\sigma$ level of confidence because of uncertainties in fitting the continuum.  Further monitoring of the water-ice absorption feature with high signal-to-noise spectra is necessary to determine if we are witnessing the processing of circumstellar ices on grains near T Tau S.

\subsection{The Infrared Variability of the T Tau Sa-Sb Binary}

Table 5 lists the cumulative results of observations that resolve T Tau Sa-Sb.  These data are drawn from our observations and from the results of Koresko (2000), Kohler, Kasper \& Herbst (2000), Duch\^ene Ghez \& McCabe (2002) and Furlan et al. (2003).   The first part of the table lists the position of T Tau Sb (the M star) with respect to Sa, and their magnitude difference [mag(Sb)-mag(Sa)].  The second part of the table lists the same information, this time for T Tau Sa with respect to N.  The table shows that both Sa and Sb vary in H and K-band flux.  T Tau Sb, the $\sim$M1 companion star, was the brighter component in observations made on 2002 Oct. 30 and 2002 Dec. 20.  In fact, in our adaptive optics observation on 2002 Oct. 30, Sa had dimmed to just 2\% of the flux of T Tau N at 2$\mu$m and was undetected in the H-band image.   This is a $\sim$20 times fainter flux level in the K-band than it was detected at in the data presented by Duch\^ene, Ghez \& McCabe (2002).  Figure 9 shows the K-band adaptive optics image that we obtained on 2002 Dec. 20 at Gemini Observatory.

Figure 10 plots the position of the unresolved T Tau S binary with respect to T Tau N using observations from 1991 to 2002.  Figures 11 and 12 show the motion of the separation and position angle of T Tau S with respect to T Tau N (see also the discussion in \S4.1).  It is apparent that the infrared variability of T Tau S arises from the variation of both components in the binary, so the apparent photocenter of T Tau S must also shift.  In Figures 10, 11 and 12 we have plotted a point that was derived from blurring the adaptive optics observation of 2002 Oct. 30 to the resolution of data obtained at the IRTF (point \#33 in Figure 10, identified by a star).  This point vividly shows the shift in photocenter, and it deviates from the linear relations plotted in the Figures 11 and 12 because T Tau Sb was the brighter component in these observations.  Much of the astrometric noise displayed in Figures 10, 11 and 12 likely arises from low level photocenter jitter.  Conversely, the photocenter jump displayed in the late 2002 data is unique in our record.  We conclude that for most of the past decade Sb has either been much fainter than T Tau Sa, or too close to it to shift the apparent position by very much.  In either case, most of the changes displayed in Figures 10, 11 and 12 have apparently traced the motion of T Tau Sa with respect to T Tau N.

\section{Orbital Motion in the T Tau System}

\subsection{The Motion of T Tau S With Respect to T Tau N}

Table 6 and Figure 10 present the apparent motion of T Tau S with respect to T Tau N from our observations described in \S2.1 and from the positions reported in the literature by Ghez et al. (1995), Roddier et al. (2000) and White \& Ghez (2001).  Overplotted in Figure 10 are identifiers that correspond to the observation number listed in Table 6.   These data show the continued motion of T Tau S with respect to T Tau N which was first reported by Ghez et al (1995).   Figures 11 and 12 show that the apparent separation and position angle have both changed from 1989 to 2002 on a pronounced level, with average motions of 2.8 milliarcseconds and 0.54$^{\circ}$ per year, respectively.  Following the discussion presented by Ghez et al. (1995), we believe that this apparent movement is best explained by orbital motion of the N-S system.  However, from these data alone it is impossible to assess their relative masses.  In the future, this could be accomplished by accurate astrometric measurements of their motion about their center of mass.  At the present time, measuring the position of T Tau N and S relative to other stars is not possible because T Tau S is not detectable in optical astrometric images, and the current infrared imaging capabilities that achieve high spatial resolution have very small fields of view.
 
\subsection{The Motion of T Tau Sb With Respect to T Tau Sa}

Figure 13 shows the apparent positions with uncertainties of T Tau Sb with respect to Sa over the course of six observations.  From the data obtained in 2002 it is clear that the movement of T Tau Sb around T Tau Sa has continued.  Following the analysis of Schaefer et al. (2003), we have fit a series of model orbits to the motion of Sb around Sa (assuming the distance to T Tau is 140 pc).  Overplotted in Figure 13 are two best-fit orbital solutions with periods of 20 and 40 years, plotted in annual intervals.  Also overplotted in Figure 13 are data points that correspond to the projected position of T Tau Sb at the time the speckle mode observations of Ghez et al. (1991) were obtained (diamonds), and the position of T Tau Sb when the lunar occultation observations of Simon et al. (1996) were made (squares).  If T Tau Sb and T Tau Sa were of comparable brightness and were at the separations delineated in Figure 13 then two components might have been detected in the high spatial resolution studies of Ghez et al. (1991), and should have been resolved by Simon et al. (1996).   In fact, the observations of Simon et al. (1996) found that T Tau S was consistent with a point source and was unresolved at a 21 mas level in the direction of the occultation.  T Tau Sb was not detected in the Ghez et al. (1991) and Simon et al. (1996) studies, so we presume that it was either too faint to be seen in speckle or lunar occultation data, or it was actually closer to T Tau Sa than the potential positions overplotted in Figure 13 and was below the resolution limit of these studies.   

We searched for orbital solutions of the T Tau Sa-Sb system by exploring the three-dimensional parameter space created by stepping through ranges of the period, time of periastron passage, and eccentricity.  For each step through the grid, the remaining four orbital elements are calculated through a linear least-squares fit (Hartkopf et al. 1989).  For orbits that are poorly determined, uncertainties in the orbital values can be derived by examining the range of parameters that produces a variation of 1 in the $\chi^2$ surface.   The orbital solutions overplotted in Figure 13 agree within $\Delta \chi^2 < 0.1$.  The possible orbital solutions that we find are derived exclusively from the relative positions of T Tau Sb with respect to Sa.  We do not use limitations on possible semi-major axes based on the knowledge that T Tau N is at a $\sim$100 AU projected distance, nor do we exclude masses for T Tau S that are inconsistent with the luminosity derived from observations.  

We performed a Monte Carlo search for solutions through periods of 10 to 1000 years, eccentricities from 0.0 to 0.99 and times of periastron passage from 1980 to 2020.  We find that the $\chi^2$ surface is broad and flat in the region of the minimum.  Therefore, it is not possible yet to determine reliable orbital parameters from the infrared data alone.  Yet, even the limited orbital data available permit estimates of the mass of the T Tau Sa/Sb binary, as described by Schaefer et al. (2003).  Figure 14 shows the distribution of possible masses that fall within the 1$\sigma$ contour interval.  There is a clear lower cutoff at 1.6 M$_{\odot}$ (1.4 M$_{\odot}$ for the 3$\sigma$ interval).  The distribution tapers off toward the high mass end, but our dynamical modeling alone can not exclude values as large as 20 M$_{\odot}$ for the total mass.   However, within the uncertainties, masses greater than $\sim$10 M$_{\odot}$ are not representative of the luminosities determined for the T Tau S system (Ghez et al. 1991; Koresko, Herbst \& Leinert 1997).  Most of these large, unphysical masses tend to come from orbits which have high eccentricities and large semimajor axes.  According to the distribution in Figure 14, for masses below 10M$_{\odot}$, the median value for the total mass of T Tau S is 4.8 M$_{\odot}$.  The spectral type of T Tau Sb is M1 (Duch\^ene, Ghez \& McCabe 2002); for a million year old star this corresponds to a mass of $\sim0.7$ M$_{\odot}$ (Baraffe et al. 1998).  Since T Tau N has been found to have a mass of $\sim$2.1 M$_{\odot}$ (White \& Ghez 2001), our orbital analysis suggests that T Tau Sa is the most massive component in the T Tau triple system.

\subsection{Comparison of Sa-Sb Orbital Motion with Radio Data}

Johnston et al. (2003) have shown convincingly that the radio emission observed toward T Tau S is not located at the same position as the infrared star.  They suggest a scenario where the radio flux arises from T Tau Sb, and the past infrared observations trace the motion of T Tau Sa.  This hypothesis is used to derive orbital parameters for the gravitationally bound T Tau Sa-Sb sytem, but this analysis does not take the bulk motion of T Tau S with respect to T Tau N into account.  Loinard et al. (2003) replotted the motion of the radio source with respect to the position of T Tau Sa using the motion of T Tau S in the infrared as discussed by Duch\^ene, Ghez \& McCabe (2002).  They suggest that T Tau Sb is in the process of being ejected from the system.  Furlan et al. (2003) obtained a high spatial resolution infrared observation of T Tau on 2002 Dec. 24 and found that the infrared position of T Tau Sb is not coincident with the projected position of the radio source in an ejection scenario.  They propose that T Tau Sb is in a bound orbit around T Tau Sa, and the radio source is a third component in the T Tau S system, T Tau Sc, which is in the process of ejection.
 
In Figures 15 and 16 we have plotted the separation and position angle of the T Tau S radio source with respect to T Tau N (Johnston et al. 2003).  Overplotted in these figures is the average linear motion determined from our infrared observations (from Figures 11 and 12).  These figures are in agreement with the conclusion from Johnston et al. (2003) that the motion of the radio source is not consistent with the infrared positions.  To rederive the motion of T Tau Sb with respect to Sa, we also assume that the radio emission arises only from T Tau Sb, and the infrared flux traces the movement of Sa (see section 3.3).  In Figure 17, all of the radio data presented by Johnston et al. (2003) are plotted with respect to the average infrared position of T Tau Sa (shown as a small cross in the center of the figure).  We have also plotted the position of the VLBI 3.6 cm radio source detected by Smith et al. (2003) in Figure 17, which they conclude is non-thermal radio emission from a kilogauss magnetic field associated with T Tau Sb.  They do not detect T Tau N in their observations, and propose that the 3.6 cm emission seen in the VLA data has been resolved out by the VLBI observation.  Using our analysis of the infrared positions of T Tau Sa and Sb, we find that the VLBI source lies within $\sim$25 mas of the position of T Tau Sa during the epoch of their observations.  It seems that this 3.6 cm radio emission may arise from extended magnetic field structure from either T Tau Sa or Sb.

In Figure 17 we have also overplotted the positions of T Tau Sb from high spatial resolution infrared observations from our study and from data available in the literature (Koresko 2000; Duch\^ene, Ghez \& McCabe 2002; Furlan et al. 2003).  The filled circles in this plot are the estimated positions of T Tau Sb derived from its position with respect to T Tau N, using the average position of T Tau Sa from Figures 11 and 12 and the same method applied to the radio data.  The filled circles and the measured positions of T Tau Sb are linked by a dashed line.  This demonstrates the inherent uncertainty of $\pm$20 mas deriving positions from the average infrared motion of T Tau Sa (overplotted as a cross in the upper left of Figure 17).  

The radio and infrared positions of T Tau Sb for the epoch 2001 observations lie within 15 mas of one another.  Thus, we find no evidence for ejection events in the T Tau system as proposed by Loinard et al. (2003) and Furlan et al. (2003).   In fact, if we use the derived positions of the 2 cm radio data in our orbital analysis (from \S4.2), we find a best-fit orbit of the Sa - Sb system which yields a period of $\sim$25 years and a total mass of T Tau S of $\sim$3.0M$_{\odot}$.  This fit is consistent with the results of Johnston et al. (2004) and is within the uncertainties of the fit presented by Tamazian (2004).  We believe that it is important to model T Tau as a triple star to get more accurate position information from the infrared versus radio data for this orbital analysis.  

The data obtained by Johnston et al. (2004) have shown that the T Tau S radio source can be resolved into two components in high spatial resolution maps.  They find that the locations of the T Tau Sa-Sb radio sources can not be associated uniquely with the positions determined from infrared observations, and the separation of the radio sources is less than the corresponding separation derived from our adaptive optics imaging.  They interpret the radio emission to arise from magnetic reconnection processes in the environment of T Tau S.   The interaction of winds, outflows and magnetic fields with the distribution of circumstellar material in the vicinity of T Tau S can likely accelerate electrons to radiate in extended clumps of non-thermal radio emission at large distances from the stars (e.g. Ray et al. 1997).  Simultaneous high spatial resolution infrared and radio monitoring of the region is necessary to further understand the nature of the radio emission and its correlation with the infrared position of T Tau Sb.






\section{T Tau N:  The Prototypical Young Star}

T Tau N is one of the most luminous and massive of the T Tauri class of young stars in the Taurus-Auriga star forming region.  It was identified as the prototypical young star in part because of its historical variability (Ambartsumian 1947; 1949; Lozinskii 1949; Beck \& Simon 2001), but in the last decade it has not varied in optical flux at a significant level (Ismaliov 1997).  We find that it also does not vary in broadband near infrared flux, but that variations in the Brackett line emission suggest fluctuations in its mass accretion rate.  T Tau N is believed to be observed nearly pole-on, with an inclination of only $\sim$8-13$^{\circ}$ with respect to the line of sight (Herbst et al. 1986).  It also has a circumstellar disk that was detected in mm continuum emission, is viewed nearly face-on, and extends $\sim$40 AU distant from the central star (Hogerheijde et al. 1997; Akeson, Koerner \& Jensen 1998).  

Muzzerole, Hartmann \& Calvet (1998) found a correlation between the Br$\gamma$ line flux and the accretion luminosity measured from the hot continuum excess emission.  They emphasize the importance of this result for classical T Tauri stars more massive than $\sim$1.5 M$_{\odot}$, because the photosperes in these young stars are hot enough to contaminate the measurement of ultraviolet continuum excess emission.  We derive estimates for the mass accretion rate from the results of our Br$\gamma$ line fluxes for T Tau N, using their Br$\gamma$ - accretion luminosity relation (Muzzerole, Hartmann \& Calvet 1998) with the virial theorem application from Gullbring et al. (1998).  We assume reasonable parameters for the inner accretion radius, approximately three to five times the radius of the star (see the analysis in Gullbring et al. 1998), and use the luminosity, mass and temperature of T Tau N from the results of White \& Ghez (2001).  The Br$\gamma$ line flux averaged over all of the seven observations of T Tau N is $\sim$6.5$\times$10$^{-16}$ W/m$^2$, which translates to an average mass accretion rate of 3.1$\times$10$^{-8}$ M$_{\odot}$/yr.  This is within 1$\sigma$ of the value derived independently by White \& Ghez (2001) based on the hot continuum excess emission, and is in the middle of the range of mass accretion rates discussed by Gullbring et al (1998).  

From the changes in Br$\gamma$ flux, we find that the mass accretion rate of T Tau N varies by a factor of four, from 1.4 to 5.9$\times$10$^{-8}$ M$_{\odot}$/yr.  This variation in accretion rate does not appear to affect the infrared K and L$'$-band flux of T Tau N at a detectable level.  Muzzerole, Hartmann \& Calvet (1998) find that both the Br$\gamma$ and Pa$\beta$ line fluxes correlate with the accretion luminosity determined from hot excess emission.  Surprisingly, we do not detect a correlation between the Br$\gamma$ and Br$\alpha$ line fluxes.   Further high spatial resolution spectroscopy of T Tau N will provide the opportunity to study the Br$\gamma$ and Br$\alpha$ line fluxes and test how their variability correlates with changes in mass accretion rate for this star.

\section{T Tau S:  The Prototypical Infrared Companion}

Although it is at a projected separation of only $\sim$100 AU from T Tau N, T Tau S is obscured completely at wavelengths shortward of 1$\mu$m.  T Tau S did not have detectable circumstellar material in the millimeter surveys of Hogerheijde et al. (1997) and Akeson, Koerner \& Jensen (1998).  However, recent HST STIS spectroscopy does provide evidence for circumbinary material enshrouding T Tau S.  Observations of the system at three position angles show a shadow at the location of T Tau S that appears to lie in front of the background ultraviolet molecular hydrogen emission around T Tau N (Walter et al. 2003). Because of the strong obscuration toward the T Tau S binary and small extinction toward T Tau N, Walter et al. (2003) describe this material as a circumbinary disk around T Tau S which is viewed nearly edge-on and is foreground to the T Tau N system.  This result supports the interpretation that the disks associated with T Tau N and S are not coplanar, as first suggested in the studies of B\"ohm \& Solf (1994) and Solf \& B\"ohm (1999) who showed that the Herbig-Haro outflows associated with T Tau (HH 155 and 255) are oriented nearly perpendicular to one another.

\subsection{The Origin of the Variability of T Tau S}

T Tau S has a pronounced IR excess emission; its large K-L$'$ color classify it as a ``high accretion rate'' star described in the study of White \& Ghez (2001).   A significant fraction of the K and L$'$-band emission from T Tau S likely arises from the reprocessing of accretion flux by the circumstellar material in the inner disk.   Hence, a cause of the extreme changes in K and L$'$-band flux might be variable accretion.  Variations in the inner disk structure that would affect the amount of circumstellar material heated by the stellar accretion flux (such as a variable inner disk radius) or changes in the mass accretion rate could be responsible for variability in the near and thermal infrared.  The calculations presented by Calvet, Hartmann, \& Strom (1997) and most of the models used in the variability study of Carpenter et al. (2001) predict that the colors of an accreting system get redder as the infrared fluxes increase, which is the opposite of what we observe.  Large color variations that have a character similar to what we have found in T Tau S have not been modeled in the context of accretion variations. 

The fact that the Br$\gamma$ emission fluctuates with the continuum suggests that these variations are linked, either by accretion or obscuration changes.  An increase in the magnetospheric accretion rate could cause an increase in the line flux, heat the inner regions of the disk ($<$ $\sim$4R$_{*}$ from the star), and thus raise the K-band continuum emission level (Gullbring et al. 1998; Calvet, Hartmann \& Strom 1997).  However, changes in the amount of obscuring material would affect the K-band continuum and Br$\gamma$ line flux in a one to one correlation, which is qualitatively similar to what is observed (Figure 7a).  When coupled with the ``redder when faint'' relation presented in Figure 5, we believe that the correlation of Br$\gamma$ and K-band flux supports the hypothesis that changes in obscuration play a dominant role in the observed flux variations in T Tau S.

If the K and L$'$-band flux variations of T Tau S are indeed caused by variable obscuration, the 2-3 magnitude amplitude suggests a fluctuation in the visible extinction of 20-30 magnitudes.  The recent decrease in K-band flux of T Tau Sa in our 2002 Oct. 30 and 2002 Dec. 20 adaptive optics observations was accompanied by a more significant decrease in H-band emission (Table 5), as expected from an increase in obscuration that follows an ISM extinction law.   This amount of visual extinction change is truly exceptional, but it may be a property of systems that contain an IRC.  Indeed, the near infrared variability of the optical component in the Haro 6-10 IRC system also exhibits K and L$'$ variability with a $''$redder when faint$''$ character similar to what we find in T Tau S (Leinert et al. 2001).

The molecular hydrogen line emission observed in T Tau S seems to vary independently from the continuum flux, though our measurements have relatively large uncertainties.  H$_{2}$ emission is not detected in the spectra of Kasper et al. (2002) and Duch\^ene, Ghez \& McCabe (2002) which sample smaller solid angles.  These results suggest that H$_2$ emission in T Tau S arises from a region that is spatially extended from the region where magnetospheric accretion flux is traced by the Brackett emission.  The H$_{2}$ emission could result from shocks in HH 255 as the outflow encounters extended circumstellar material in the vicinity of T Tau S.  

Since the discovery of T Tau S (Dyck, Simon \& Zuckerman 1982), a number of authors have reported additional components in the T Tau system.  In the hope of detecting T Tau S in the visible, Nisenson et al. (1985) obtained speckle images and discovered an optical source 0.$''$27 North of T Tau N (PA $\sim$ 358$^{\circ}$).  A few years later, Mahaira \& Kataza also detected a source at optical and near infrared wavelengths at a separation of 0.$''$40 North of T Tau N.  In the radio, Ray et al. (1997) discovered another companion 0.$''$29 to the south of T Tau N (PA $\sim$195$^{\circ}$).  The optical sources have since disappeared, they were not detected in the deep imaging studies of Gorham et al. (1995) or Stapelfeldt et al. (1998).  Additionally, the third radio source discovered in the study of Ray et al. (1997) has not been reported in the observations of Johnston et al. (2003).  To date, a satisfactory explanation of the appearance and disappearance of such ``companions'' in the T Tau system has not been made.   However, the positions of each of these sources are to the north or south of T Tau N, and all of them lie coincidentally along the axis of the redshifted component of the HH 255 outflow from T Tau S (Solf \& B\"ohm 1999).  Several of the companions detected in the past studies of the T Tau system may have been optical emission clumps or radio knots associated with the HH 255 outflow from the T Tau S binary.

\subsection{The Nature of T Tau S}

If T Tau Sa is a 3-4 M$_{\odot}$ young star with an age of $\sim$1 Myr, as our results and the model of Koresko, Herbst \& Leinert (1997) indicate, its apparent magnitude at V would be around $\sim$6 mag (Siess, Dufour \& Forestini 2000; assuming a 140 pc distance).  It would be intrinsically brighter than T Tau N at optical and near infrared wavelengths and would be well on its way to contracting onto the main sequence to become an early A or late B-type star.  An early spectral type for T Tau Sa could explain why no photospheric features typical of young low mass stars are deteced in its K-band spectra.  In this scenario, more than 10 magnitudes of visual extinction would be required for T Tau S to escape detection in the optical study of Stapelfeldt et al. (1998).  

The water-ice and silicate signatures observed toward T Tau S do not imply obscurations as large the $\sim$35 magnitudes required to obscure it in the model presented by Koresko, Herbst \& Leinert (1997).  However, the  A$_v$ values derived from the observations are dependent on which optical depth versus visual extinction relation is used, and the applicability of these relations toward very obscured young stars is uncertain (Beck, Prato \& Simon 2001; Herbst Robberto \& Beckwith 1997; Ghez et al. 1991, Teixeira \& Emerson 1999).   Warm dust would not show water-ice absorption if the temperature was high enough that all of the ices along the line of sight had sublimated, hence water-ice is not the best tracer of all the material that obscures T Tau S.  If there are 35 magnitudes of extinction along the line of sight toward T Tau S then it should have a deep silicate absorption feature.  The fact that the silicate absorption does not indicate the presence of 35 magnitudes of optical obscuration in the studies of Ghez et al. (1991) and Herbst, Robberto \& Beckwith (1997) could be a result of grain processing within the circumstellar material, and the A$_v$/$\tau_{silicate}$ ratio may be inherently different than the ISM value which has been used to interpret the observations. 

It seems plausible that some circumstellar material in the vicinity of T Tau Sa is distributed in a nearly edge-on disk that results in significant obscuration at optical wavelengths, and that this material was seen as the shadow in the spectral study of Walter et al. (2003).  Young stars with circumstellar disks viewed edge-on have been discovered through the light scattered off the surface of their disks (Padgett et al. 1998; Koresko 1998).  Results from the modeling of these systems suggest that the optically thick disk midplanes can have hundreds or thousands of magnitudes of visual extinction through their extent (Stapelfeldt et al. 1998b; Wolf, Padgett \& Stapelfeldt 2003).   However, the triple nature of T Tau complicates the spectral energy distribution model for T Tau S that was presented in Koresko, Herbst \& Leinert (1997).   

Theoretical studies on the stability of circumstellar material in binary systems predict that disks around the stars are truncated at a fraction of the semi-major axis of the orbit.  The maximum possible disk extent is determined from the mass ratio, eccentricity and separation at periastron (Artymowicz \& Lubow 1994).   Although we are able to derive an estimate for the T Tau S system mass from the distribution plot presented in Figure 14, we are unable to determine unique values for the eccentricity and periastron distance from our analysis.  However, the orbital model presented by Johnston et al. (2004) estimates the periastron distance at $\sim$5 -6 AU, hence a circumstellar disk encircling T Tau Sa would need to have an extent less than this.  Detailed  modeling of the spectral energy distribution is necessary to determine if it is possible to extinguish the light from an edge-on disk with a $\sim$6AU extent around T Tau Sa in order to reproduce the observations. 




\section{Summary}

\subsection{Results for T Tau N:}

1)  T Tau N does not vary at K and L$'$ on a statistically significant level.  In the interval we considered, its average K and L$'$ brightness was 5.53$\pm$0.03 and 4.32$\pm$0.05 magnitudes, respectively.

2)  We find no evidence for the 3 $\mu$m ice-band feature in absorption, consistent with the relatively small, A$_v\sim$1.5 magnitude extinction measured toward T Tau N.

3)  Br$\gamma$ and Br$\alpha$ line emission indicates that T Tau N is actively accreting material from its circumstellar environment, and we detected photospheric absorption features of Na, and Ca, and CO consistent with its K0 spectral type. 

4)  The emission of atomic hydrogen varies by nearly a factor of four; the average value suggests an underlying mass accretion rate of 3.1$\times$10$^{-8}$ M$_{\odot}$/year.  This accretion variability does not have an affect on the K and L$'$-band fluxes of the star.

\subsection{Results for T Tau S (the IRC):}

1)  The near infrared flux of T Tau S varies significantly in both the K and L$'$ photometric bands on timescales as short as a week.

2)  The K-L$'$ versus K-band magnitude shows a clear correlation which suggests that one process dominates the variability observed in T Tau S.  The color-magnitude relation shows a ``redder when faint'' character that is not well explained by current accretion models, but is very similar to the relation that would be expected if variable extinction along the line of sight is partly responsible.  The range of magnitude variations observed in the K and L$'$ bands suggest unusually large variations in obscuring material.

3)  T Tau S is always more obscured than T Tau N, as indicated by its redder color and the absorption in the water-ice feature at 3$\mu$m.  The ice-band optical depth varies by a factor of 2-3, again suggesting that variations in the amount of obscuring material along the line of sight may play a large role in the variability of T Tau S.

4)  We find the Br$\gamma$ and Br$\alpha$ lines in emission in the spectra of T Tau S and find that their emission varies by factors of 3-4.

5)  We confirm the detection of the S(1) {\it v}=1-0 H$_{2}$ feature in emission toward T Tau S; there is no evidence for this feature in the spectra of T Tau N.  We find no emission from other molecular hydrogen features observable in the 2.0 - 2.45 $\mu$m wavelength region.  The S(1) {\it v}=1-0 H$_{2}$ emission varies by a factor of $\sim$2.  

6) We present orbital data for the unresolved T Tau S system with respect to T Tau N, and for T Tau Sb with respect to T Tau.  We find that T Tau Sa and Sb are probably bound and that T Tau Sa is likely the most massive star in the T Tau triple system.

7)  We have used the motion that we derive for T Tau S in the infrared to replot the observed VLA positions of the T Tau S radio source.  The epoch 2001 radio and infrared positions of T Tau Sb are consistent to within a 1$\sigma$ uncertainty.  We find no evidence for an ejection within the system.

8)  Our interpretation is that T Tau Sa is likely a young star that is observed through a significant and variable amount of obscuration and is actively accreting material from its circumstellar environment.  Further modeling is necessary to determine if material in an edge-on circumstellar disk within the orbit of the Sa-Sb binary can extinguish the light from T Tau Sa to obscure it to invisibility at optical wavelengths.

\acknowledgments

We thank Andrea Ghez for helping to initiate this project by contributing unpublished photometry from 1994-1995, and Markus Kasper for sharing his unpublished measurements from 2000.  Our friends at the NASA Infrared Telecope Facility, David Griep, Bill Golish, Bobby Bus, Paul Sears, Alan Tokunaga and Charlie Kaminski made this project possible by carrying out service observations and assisting us with remote observing.    Mike Cushing and John Rayner provided valuable advice in data reduction using the SpeXTool reduction software.  We also thank our anonymous referee for carefully reading our manuscript and providing us with suggestions for improving it.  We are indebted to the Altair adaptive optics team for observing T Tau as a commissioning target.  The work of GS and MS was supported by NSF Grant 02-05427.     This study was supported in part by the Gemini Observatory, which is operated by the Association of Universities for Research in Astronomy, Inc., on behalf of the international Gemini partnership of Argentina, Australia, Brazil, Canada, Chile, the United Kingdom, and the United States of America.  The authors extend a special thanks to those of Hawaiian ancestry on whose sacred mountain we are privileged to be guests.



\appendix




\clearpage










\clearpage


\clearpage
\begin{deluxetable}{lcccc}
\tabletypesize{\scriptsize}
\tablecaption {Log of the Observations of T Tau \label{tbl-1}}
\tablewidth{0pt}
\tablehead{
\colhead{UT Date} & \colhead{Result}   & \colhead{Telescope}   &
\colhead{Instrument}  & \colhead{Note} 
}
\startdata
08 Aug 1995   & K, L$'$ & IRTF & NSFCam & \\ 
15 Sep 1995   & L$'$ & IRTF & NSFCam & \\
27 Oct 1996   & K, L$'$ & IRTF & NSFCam & \\ 
13 Dec 1997 & K, L$'$ & IRTF & NSFCam & 1  \\
6 Jan 1998 & K, L$'$ & IRTF & NSFCam & 1  \\ 
16 Jan 1998 & K, L$'$ & IRTF & NSFCam & 1 \\  
18 Jan 1998 & K, L$'$ & IRTF & NSFCam & 1 \\  
8 Feb 1998  & K, L$'$ & IRTF & NSFCam & 1 \\ 
6 Mar 1998 & K, L$'$ & IRTF & NSFCam & \\ 
7 Mar 1998  & K, L$'$ & IRTF & NSFCam & \\ 
9 Mar 1998  & K, L$'$ & IRTF & NSFCam & 1 \\ 
13 Sep 1998  & K, L$'$ & IRTF & NSFCam & \\ 
14 Sep 1998  & K, L$'$ & IRTF & NSFCam &  \\ 
15 Sep 1998  & K, L$'$ & IRTF & NSFCam &  \\ 
09 Dec 1998 & K, L$'$ & IRTF & NSFCam & \\
11 Dec 1998 & K, CVF filters$^*$, L$'$ & IRTF & NSFCam & \\ 
24 Feb 1999 & K, L$'$ & IRTF & NSFCam & \\ 
17 Sep 1999  & K$_{sys}$, L$'$$_{sys}$ & WIRO & IoCam 1 & \\  
19 Sep 1999 & K$_{sys}$, L$'$$_{sys}$ & WIRO & IoCam 1 &  \\ 
22 Sep 1999 & K$_{sys}$, L$'$$_{sys}$ & WIRO & IoCam 1 & \\ 
23 Sep 1999 & K$_{sys}$, L$'$$_{sys}$ & WIRO & IoCam 1 & \\ 
6 Oct 1999 & K$_{sys}$, L$'$$_{sys}$ & WIRO & IoCam 1 & \\ 
1 Nov 1999  & K, L$'$ & IRTF & NSFCam & \\  
2 Nov 1999  & K, L$'$ & IRTF & NSFCam & \\  
3 Nov 1999  & K, L$'$ & IRTF & NSFCam & \\  
17 Nov 1999  & K, CVF filters$^*$, L$'$ & IRTF & NSFCam & 1 \\  
27 Jan 2000  & K, CVF filters$^*$, L$'$ & IRTF & NSFCam & 1 \\  
10 Oct 2000 & K, L$'$, 2-4$\mu$m spectra & IRTF & SpeX & \\
11 Nov 2000 & K, 2-4$\mu$m spectra & IRTF & SpeX & 2 \\
18 Nov 2000 & 2-4$\mu$m spectra & IRTF & SpeX & \\
12 Jan 2001 & K, L$'$ & IRTF & NSFCam & \\
15 Oct 2001 & K, L$'$, 2-4$\mu$m spectra & IRTF & SpeX & \\ 
10 Nov 2001 & K, L$'$ & IRTF & NSFCam & \\ 
2 Jan 2002 & K, L$'$, 2-4$\mu$m spectra & IRTF & SpeX & \\ 
30 Oct 2002 & H, K$'$, Br$\gamma$ & Keck & NIRC2+Keck AO \\
20 Dec 2002 & H, K, H$_2$  & Gemini & NIRI+Altair AO \\
\enddata
\tablecomments{A log of the observations made for this project. Notes: 1 - Observations made in Service Mode.  2 - Observations made remotely from SUNY Stony Brook.  $^*$-CVF Filter data presented in Beck, Prato \& Simon 2001. The  K$_{sys}$ and L$'$$_{sys}$ designations used for data obtained at WIRO refer to spatially unresolved observations of the total N-S system flux.}
\end{deluxetable}

\clearpage




\clearpage

\begin{deluxetable}{llccccc}
\tabletypesize{\scriptsize}
\tablecaption {The Infrared Photometric Variability of T Tau \label{tbl-2}}
\tablewidth{0pt}
\tablehead{
\colhead{Obs. Number} & \colhead{Date (UT)}   & \colhead{K$_{TTN}$}   &
\colhead{K$_{TTS}$}  & \colhead{L$'_{TTN}$} & \colhead{L$'_{TTS}$} &
\colhead{Reference} 
}
\startdata

1 & 13 Oct 1981	 & 5.6	 & 8.4	 & 4.3 & 5.6$^{a}$ & Dyck, Simon \& Zuckerman (1982) \\ 
2 & 1 Dec 1983 & -- & 8.4 & -- & -- & Beckwith et al. (1984)  \\ 
3 & 16 Dec 1988 & --  & 8.5$^{b}$ & -- & -- & Maihara \& Kataza (1991)\\ 
4 & 16 Aug 1989 & 5.6  & 7.6 & -- & -- & Ghez et al. (1991) \\ 
5 & 9 Dec 1989 & 5.6  & 6.7 & -- & -- & Ghez et al. (1991) \\       
6 & 7 Aug 1990 & 5.6  & 6.5 & -- & -- & Ghez et al. (1991) \\       
7 & 3 Oct 1990 & --  & -- & 4.6	& 3.7 & Ghez et al. (1991) \\       
8 & 9 Nov 1990 & 5.6  & 6.5 & -- & -- & Ghez et al. (1991) \\       
9 & 4 Dec 1990 & --  & -- & -- & 3.5$^{a}$ & Tessier, Bouvier \& Lacombe (1994) \\
10 & 24 Jan 1992 & 5.5  & 7.6 & -- & -- & Kobayashi et al. (1994) \\
11 & 20 Oct 1994 & 5.6 & 8.4 & 4.5 & 4.8 & A. Ghez (private comm.) \\


12 & 16 Dec 1994 & 5.5  & 7.8 & 4.3 & 4.6 & Simon et al. (1996) \\
13 & 25 Dec 1994 & --  & 8.2 & -- & -- & Roddier et al. (2000) \\

14 & 13 Apr 1995 & -- & 8.5 & -- & 5.0 & A. Ghez (private comm.) \\


15 & 08 Aug 1995 & 5.5 & 7.6 & 4.4 & 4.6 & \\
16 & 15 Sep 1995 & --  & --  & --  & 4.5 & \\
17 & 15 Nov 1995 & 5.5 & 7.8 & 4.5 & 4.7 & A. Ghez (private comm.) \\


18 & 31 Aug 1996 &  5.3 & 6.7 & -- & -- & Herbst et al. (1997) \\
19 & 23 Oct 1996 & -- & 7.5$^{b}$ & -- & -- & Roddier et al. (2000) \\
20 & 27 Oct 1996 &  5.5 & 7.8 & 4.4 & 4.2 \\
21 & 17 Nov 1997 & --  & 7.9$^{b}$ & -- & -- & Roddier et al. (2000) \\
22 & 04 Dec 1997 & --  & 7.4$^{b}$ & -- & -- & White \& Ghez (2001) \\
23 &  06 Dec 1997 & -- & 7.4$^{b}$ & -- & 4.0$^{b}$ & White \& Ghez (2001) \\
24 &  13 Dec 1997 & 5.6 & 6.5 & 4.3 & 4.2  & \\  
25 &  15 Dec 1997 & 5.7 & 6.7 & -- & --  & Koresko (2000)\\  
26 &  6 Jan 1998 & 5.5 & 7.0 & 4.3 & 3.9   & \\ 
27 &  16 Jan 1998 & 5.5 & 6.8 & 4.3 & 4.1  & \\  
28 &  18 Jan 1998 & 5.6 & 6.7 & 4.4 & 4.0  & \\  
29 &  8 Feb 1998 & 5.6 & 6.1 & 4.3 & 3.6  & \\ 
30 &  6 Mar 1998 & 5.6 & 6.2 & 4.3 & 3.7  & \\ 
31 &  7 Mar 1998 & 5.6 & 6.3 & 4.4 & 4.0  & \\ 
32 &  10 Mar 1998  & 5.5 & 6.5 & 4.2 & 3.9  & \\ 
33 &  13 Sep 1998 & 5.6 & 6.7 & 4.4 & 4.0  & \\
34 &  14 Sep 1998 & 5.6 & 6.8 & 4.4  & 4.0    & \\ 
35 &  15 Sep 1998  & 5.6 & 7.0 & 4.4 & 4.1  & \\ 
36 &  2 Nov 1998 & --  & 7.0$^{b}$ & -- & -- & Roddier et al. (2000) \\
37 &  09 Dec 1998  & --  & 6.8 & --  & 4.1 & \\
38 &  11 Dec 1998  & 5.6 & 7.0 & 4.4 & 4.2  & \\
39 &  24 Feb 1999  & --  & 6.9 & --  & 4.2  & \\  
40 &  17 Sep 1999  & --  & 5.8 & --  & 3.4  & \\  
41 &  19 Sep 1999  & --  & 5.9 & --  & 3.4   & \\ 
42 &  22 Sep 1999  & --  & 6.0 & --  & 3.5  & \\ 
43 &  23 Sep 1999  & --  & 5.9 & --  & 3.6  & \\ 
44 &  28 Sep 1999  & 5.3  & 6.2 & --  & --  & Kasper et al. (2002) \\ 
45 &  6 Oct 1999  & --  & 6.1 & --  & 3.6  & \\ 
46 & 1 Nov 1999 & -- & 6.2$^{b}$ &  --  & 4.1$^{b}$ &  \\  
47 & 2 Nov 1999 & -- & 6.2$^{b}$ & -- & 4.0$^{b}$ & \\
48 & 3 Nov 1999 & 5.4 & 6.1 & 4.3 & 3.9 & \\
49 & 17 Nov 1999  & 5.6 & 6.2 &  4.5 &  4.1 &  \\
50 & 23 Nov 1999 & --  & 6.2$^{b}$ & -- & --  &  Roddier et al. (2000)    \\
51 &  27 Jan 2000  & 5.5 & 6.2 &  4.4 & 3.9  & \\  
52 &  20 Feb 2000  & 5.5  & 6.7 & -- & -- & M. Kasper (private comm.)\\ 
53 &  10 Oct 2000  & 5.6 & 6.6 & 4.3 & 4.1   & \\ 
54 &  11 Nov 2000  & 5.5 &  6.4  & --  & 3.7$^{b}$ & \\ 
55 &  18 Nov 2000  & --  &  6.4$^{b}$ & -- & -- & \\
56 &  19 Nov 2000  & -- &  6.3 & --  & -- &  Duch\^ene, Ghez \& McCabe (2002)\\ 
57 &  12 Jan 2001  & 5.5 &  6.9 & 4.4 & 4.1  & \\
58 &  15 Oct 2001  & 5.5 &  6.2 & 4.3  & 4.1 & \\ 
59 &  10 Nov 2001  & -- &   7.0 & --  & 4.5 & \\
60 &  2 Jan 2002  & 5.5 &  7.3 & 4.4  & 4.7 & \\ 
61 &  30 Oct 2002 & -- & 8.2 & -- & -- & \\
62 &  24 Dec 2002 & 5.6 & 7.5 & -- & -- & Furlan et al. (2003) \\
\enddata
\tablecomments{$^{a}$ - Observations presented in Dyck, Simon \& Zuckerman and Tessier, Bouvier \& Lacombe were obtained in the L-band and may be 0.1 to 0.15 magnitudes fainter than the values we obtain in the L$'$-band.  $^{b}$ - Magnitude of T Tau S is determined using the average magnitude of T Tau N combined with:  1) flux ratios derived from our imaging and spectral observations, 2) the flux ratio of the T Tau N/S system presented in the literature or 3) The sum of the flux of the T Tau Sa-Sb system.  The typical uncertainties in the magnitudes are $\sim\pm$0.1.}
\end{deluxetable}

\clearpage

\begin{deluxetable}{lcccccc}
\tabletypesize{\scriptsize}
\tablecaption {Spectral Line Variability in T Tau \label{tbl-3}}
\tablewidth{0pt}
\tablehead{
\colhead{Obs. \#} & \colhead{T Tau N Br$\gamma$} & 
\colhead{T Tau N Br$\alpha$} & \colhead{T Tau N H$_{2}$} & 
\colhead{T Tau S Br$\gamma$} & \colhead{T Tau S Br$\alpha$} & 
\colhead{T Tau S H$_{2}$} \\
 & (10$^{-16}$ W/m$^{2}$) & (10$^{-16}$ W/m$^{2}$) & (10$^{-16}$ W/m$^{2}$) & (10$^{-16}$ W/m$^{2}$) & (10$^{-16}$ W/m$^{2}$) & (10$^{-16}$ W/m$^{2}$) \\ 
}
\startdata
20 & 8.3$\pm$0.3  & -- & $<$0.9 (3$\sigma$) & 1.4$\pm$0.3 & -- & 0.6$\pm$0.3 \\
44 & 9.7$\pm$1.1  & -- & $<$0.9   & 2.7$\pm$0.3 & -- & $<$0.9 (3$\sigma$) \\
53 & 6.2$\pm$0.3  & 3.4$\pm$0.6 & $<$0.6 & 2.8$\pm$0.3 & 3.7$\pm$0.5 & 0.7$\pm$0.2 \\
54 & 4.5$\pm$0.3  & 5.6$\pm$0.5 & $<$0.9 & 3.1$\pm$0.3 & 7.7$\pm$0.6 & 0.4$\pm$0.3 \\
55 & 3.5$\pm$0.3  & -- & $<$0.6 & 4.2$\pm$0.3 & --  & 0.3$\pm$0.2 \\
58 & 4.8$\pm$0.4  & 1.5$\pm$0.8 & $<$1.2 & 4.0$\pm$0.3 & 4.3$\pm$0.8  & 1.0$\pm$0.4 \\
60 & 8.9$\pm$0.2  & 5.9$\pm$0.6 & $<$0.6 & 2.3$\pm$0.2 & 3.0$\pm$0.5  & 0.9$\pm$0.2 \\
\enddata
\end{deluxetable}

\clearpage

\begin{deluxetable}{lcccccc}
\tabletypesize{\scriptsize}
\tablecaption {Variability in Water-ice Absorption toward T Tau S  \label{tbl-4}}
\tablewidth{0pt}
\tablehead{
\colhead{Obs. \#} & \colhead{T$_{BB}$} & 
\colhead{$\tau_{ice}$} & \colhead{FWHM} & 
\colhead{N$_{ice}$} & \colhead{A$_v$ (TE99)} & 
\colhead{A$_v$ (W88)} \\
 & &  &  cm$^{-1}$ & $\times 10^{18}$ cm$^{-2}$ & & \\ 
}
\startdata
38$^{*}$ &    --  & 0.5$\pm$0.1  &   --     &      --     &       --     &    9$\pm$2 \\
49$^{*}$ &    --  & 0.3$\pm$0.1   &   --     &      --     &       --    &     7$\pm$2 \\
51$^{*}$ &    --  & 0.2$\pm$0.1   &   --      &     --     &      --     &    5$\pm$2 \\
53  &   890 & 0.65$\pm$0.09  &  460$\pm$20 & 1.5$\pm$0.3 & 17$\pm$4 &  10$\pm$1 \\ 
54  &   840 & 0.7$\pm$0.2   &  520$\pm$30 & 1.8$\pm$0.5 & 20$\pm$5 & 11$\pm$2 \\
55  &   870 & 0.58$\pm$0.08    &   480$\pm$20  &  1.4$\pm$0.3   &   16$\pm$4 & 9.6$\pm$0.8  \\
58  &   1320 &  0.96$\pm$0.09   &    510$\pm$20  &  2.5$\pm$0.4 &  27$\pm$7 & 13$\pm$1 \\
60  &   1030 & 0.68$\pm$0.08   &    720$\pm$30 & 2.5$\pm$0.5  &   27$\pm$6 & 11$\pm$1 \\
\enddata
\tablecomments{$^{*}$ Water-ice data from circular variable filter (CVF) observations are presented in Beck, Prato \& Simon (2001).  TE99 refers to visual extinctions derived using the relation discussed in Teixeira \& Emerson (1999) and W88 values were derived using the $\tau_{ice}$ versus A$_{v}$ correlation found by Whittet et al. (1988).}
\end{deluxetable}

\clearpage

\begin{deluxetable}{lccccc}
\tabletypesize{\scriptsize}
\tablecaption {Apparent Motion and Variability of the T Tau S binary\label{tbl-6}}
\tablewidth{0pt}
\tablehead{
\colhead{UT Date}  & \colhead{Position Angle}  &
\colhead{Separation}  & \colhead{$\Delta$K}  & \colhead{$\Delta$H} &
\colhead{Reference} 
}
\startdata
\multicolumn {5}{l} {Sb with respect to Sa:} \\
15 Dec. 1997   &  225$\pm$8 &  0.$''$053$\pm$0.$''$009 & 2.6 & -- & Koresko (2000) \\
20 Feb. 2000 &   253$\pm$2 &  0.$''$079$\pm$0.$''$002 & --  & 1.5 & Kohler, Kasper \& Herbst (2000) \\
19 Nov. 2000 & 267$\pm$2 & 0.$''$092$\pm$0.$''$003 & 1.89 & 1.17 & Duch\^ene, Ghez \& McCabe (2002) \\
30 Oct. 2002 &  283.4$\pm$2.1 & 0.$''$107$\pm$0.$''$004 & -1.36 & $>$-3.7 & \\
20 Dec. 2002 &  282.0$\pm$2.1 & 0.$''$110$\pm$0.$''$004 & -0.9 & -2.37 & \\
24 Dec. 2002 &  287.8$\pm$2.6 & 0.$''$108$\pm$0.$''$005 & 0.0$^{*}$ & -- & Furlan et al. 2003 \\
\hline
\multicolumn {5}{l} {Sa with respect to N:} \\
15 Dec. 1997  &  --  &  -- & 1.3 & -- & Koresko (2000) \\
19 Nov. 2000 & 179.7$\pm$0.2 & 0.$''$702$\pm$0.$''$005 & 1.36 & 3.17 & Duch\^ene, Ghez \& McCabe (2002) \\
30 Oct. 2002  &  183.1$\pm$0.3 & 0.$''$697$\pm$0.$''$002 & 4.24 & $>$7.5 & \\
24 Dec. 2002  &   181.7$\pm$0.4 & 0.$''$691$\pm$0.$''$002 & 2.6$^{*}$ & -- & Furlan et al. 2003 \\ 
\enddata
\tablecomments{$^{*}$ - Observations from Furlan et al. (2003) were made in the 2.26$\mu$m narrow-band filter.  Based estimates from our past narrow band and spectral observations, the flux ratios derived from these data should be very similar to K-band observations ($<$5\% deviation).  Positions of T Tau Sa with respect to N are not presented for 20 Feb. 2000 and 20 Dec. 2002 because Kohler, Kasper \& Herbst (2000) did not include this information in their study, and because T Tau N was saturated in the adaptive optics images obtained at Gemini. }
\end{deluxetable}
\clearpage

\begin{deluxetable}{llccc}
\tabletypesize{\scriptsize}
\tablecaption {Apparent motion of T Tau S with respect to T Tau N\label{tbl-5}}
\tablewidth{0pt}
\tablehead{
\colhead{Orbital Obs.} & \colhead{Date} & \colhead{Position Angle}  &
\colhead{Separation}  & \colhead{Reference} \\
}
\startdata
1 & 1989 Dec 10 &  175.5$\pm$ 0.2 &  0.720$\pm$0.004  & Ghez et al. (1995) \\
2 & 1990 Nov 10 &  176.15$\pm$0.02 & 0.7155$\pm$0.0005 & Ghez et al. (1995) \\
3 & 1991 Nov 19 &  176.41$\pm$0.05 &  0.702$\pm$0.003 & Ghez et al. (1995) \\
4 & 1992 Feb 20 &  176.69$\pm$0.09 &  0.710$\pm$0.004 & Ghez et al. (1995) \\
5 & 1992 Oct 11 &  177$\pm$1 &  0.692$\pm$0.008 & Ghez et al. (1995) \\
6 & 1993 Nov 25 &  177.0$\pm$0.2 &  0.701$\pm$0.004 & Ghez et al. (1995)  \\
7 & 1993 Dec 26 &  176.0$\pm$0.5 &  0.690$\pm$0.005 & Ghez et al. (1995) \\
8 & 1994 Sep 22 &  175.9$\pm$0.6 &  0.689$\pm$0.005 & Ghez et al. (1995) \\
9 & 1994 Nov 19 &  178.2 $\pm$0.4 &  0.689$\pm$0.006 & Ghez et al. (1995) \\
10 & 1994 Dec 25 &  177.6$\pm$0.6 &  0.718$\pm$0.004 &  Roddier et al. (2000) \\
11 & 1996 Oct 27 &  179.4$\pm$0.2 & 0.708$\pm$0.007 &  \\
12 & 1997 Nov 17 &  180.2$\pm$0.7 &  0.709$\pm$0.008 &  Roddier et al. (2000) \\
13 & 1997 Dec 04 &  179.5$\pm$1.0 & 0.685$\pm$0.013  &  White \& Ghez (2001) \\
14 & 1997 Dec 06 &  179.1$\pm$1.0 & 0.698$\pm$0.013  &  White \& Ghez (2001) \\
15 & 1997 Dec 13 &  179.8$\pm$0.5 & 0.72$\pm$0.02 & \\
16 & 1998 Jan 06 &  179.7$\pm$0.4 & 0.711$\pm$0.005 & \\ 
17 & 1998 Jan 16 &  179.2$\pm$0.7 & 0.714$\pm$0.008 & \\
18 & 1998 Jan 18 &  179.5$\pm$0.5 & 0.711$\pm$0.007 & \\
19 & 1998 Feb 08 &  179.6$\pm$0.4 & 0.713$\pm$0.006 & \\ 
20 & 1998 Mar 06 &  179.7$\pm$0.4 & 0.703$\pm$0.006 & \\
21 & 1998 Mar 07 &  179.9$\pm$0.4 & 0.709$\pm$0.005 & \\
22 & 1998 Mar 08 &  179.4$\pm$0.7 & 0.708$\pm$0.008 & \\
23 & 1998 Sep 13 &  179.8$\pm$0.4 & 0.714$\pm$0.006 & \\ 
24 & 1998 Nov 02 &  180.7$\pm$0.9 &  0.704$\pm$0.007 &  Roddier et al. (2000) \\
25 & 1998 Dec 09 &  179.8$\pm$0.4 & 0.69$\pm$0.01 & \\
26 & 1998 Dec 11 &  180.0$\pm$0.5 & 0.701$\pm$0.005 & \\
27 & 1999 Feb 24 &  180.0$\pm$0.4 & 0.688$\pm$0.006 & \\ 
28 & 1999 Nov 1  & 180.5$\pm$0.5  & 0.67$\pm$0.01 & \\
29 & 1999 Nov 18 &  180.5$\pm$0.5 & 0.701$\pm$0.006 &  Roddier et al. (2000) \\
30 & 1999 Nov 23 &  181.0$\pm$0.1 &  0.699$\pm$0.006 & \\
31 & 2001 Jan 16 &  180.9$\pm$0.5 & 0.695$\pm$0.006 & \\
32 & 2001 Nov 10 &  181.5$\pm$0.4 & 0.677$\pm$0.005 & \\
33 & 2002 Oct 30 &  191.3$\pm$0.2 & 0.689$\pm$0.002 & Convolved AO Data \\
\enddata
\end{deluxetable}

\clearpage



\begin{figure}
\epsscale{1.0}
\includegraphics[angle=90, width=17.5cm, height=13.0cm]{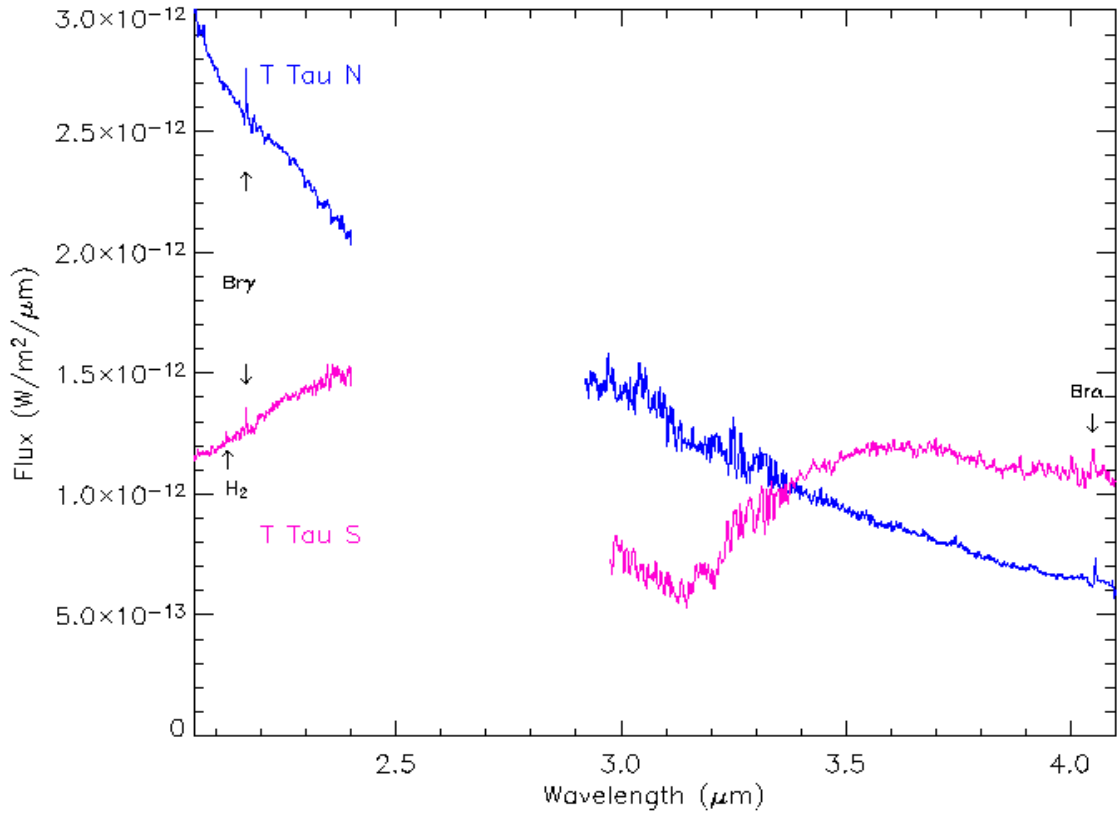}
\caption{The 2-4$\mu$m spectra of T Tau N and S measured on 2001 Oct. 15 using SpeX.  The Hydrogen Br$\gamma$ and Br$\alpha$ features can be seen at 2.16 and 4.05 $\mu$m in both spectra while molecular hydrogen emission (2.12 $\mu$m) and the deep, broad 3.05 $\mu$m absorption feature of water-ice on dust grains are seen only in the spectra of T Tau S. \label{fig1}}
\end{figure}
\clearpage

\begin{figure}
\epsscale{0.8}
\includegraphics[angle=90, width=17.5cm, height=13.0cm]{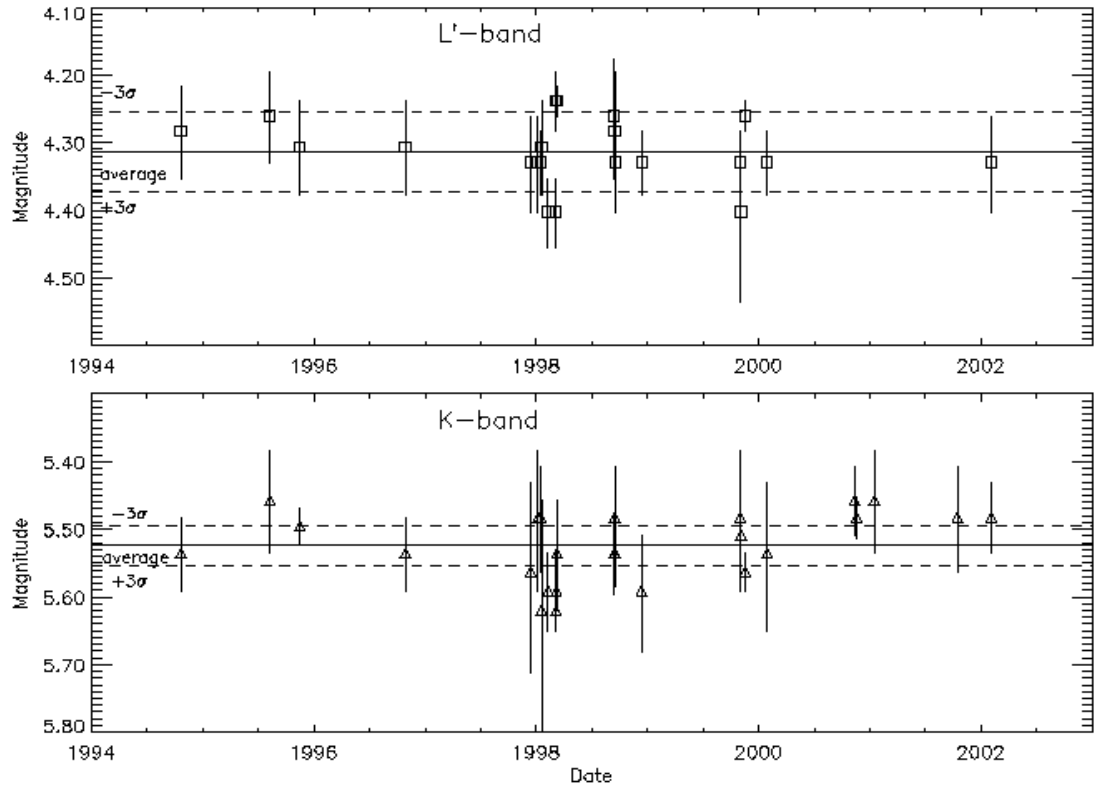}
\caption{The variability of T Tau N in the K and L$'$-bands plotted from our observations made during photometric conditions.  T Tau N has not varied at a statistically significant level during the course of our monitoring campaign.  The average magnitudes are 5.53$\pm$0.03 in the K-band and 4.32$\pm$0.05 at L$'$.\label{fig2}}
\end{figure}
\clearpage

\begin{figure}
\epsscale{0.8}
\includegraphics[angle=90, width=17.5cm, height=13.0cm]{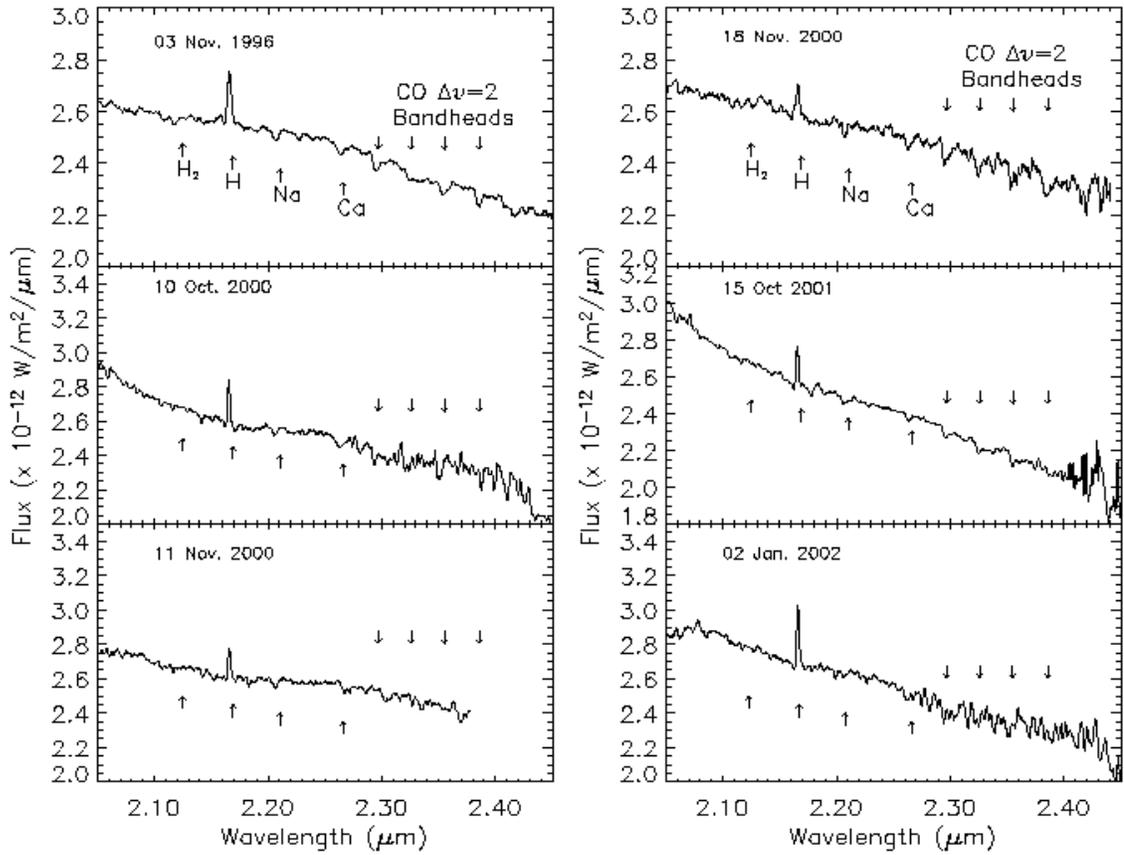}
\caption{The K-band (2.0-2.45 $\mu$m) spectra of T Tau N.  Emission of Hydrogen Br$\gamma$ is detected in all of the spectra, and the photospheric absorption features of Na, Ca, and the CO $\Delta${\it v}=2 bandheads are seen in several of the spectra. \label{fig3}}
\end{figure}
\clearpage

\begin{figure}
\epsscale{0.8}
\includegraphics[angle=90, width=17.5cm, height=13.0cm]{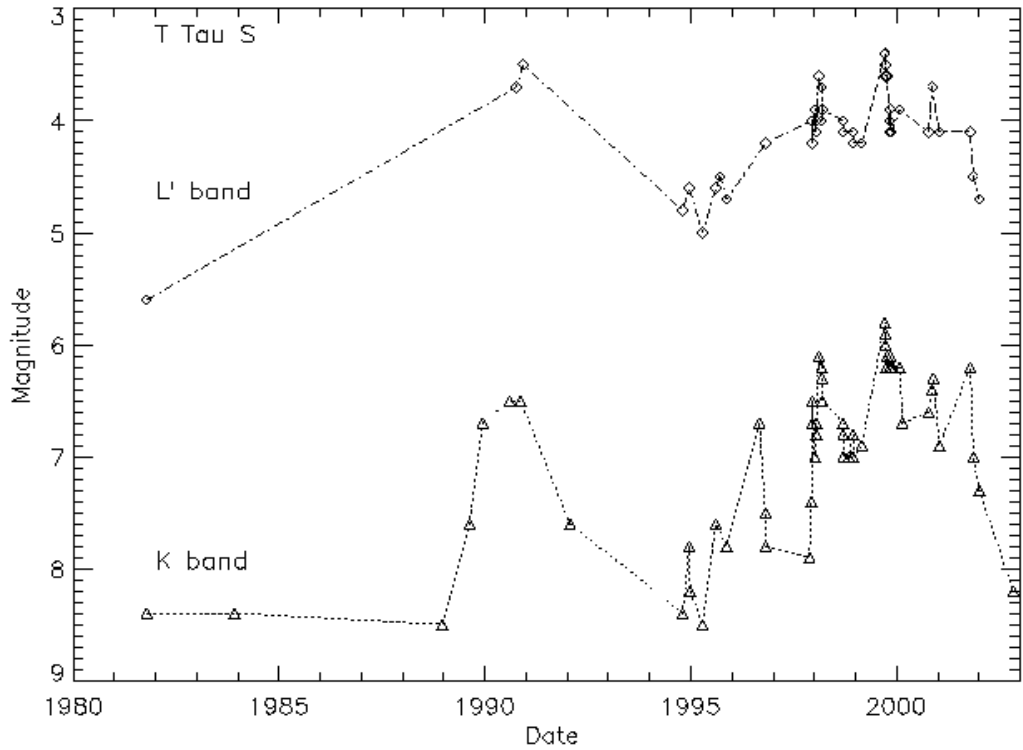}
\caption{The K and L$'$-band infrared variability of T Tau S from all spatially resolved observations made since its discovery in 1981. The K-band magnitude of T Tau S has varied by $\sim$3, and the L$'$-band has varied by more than 2 magnitudes.  The lines overplotted in this figure are meant to aid the eye and not to imply continuity of the brightness variations.  The typical uncertainties in the magnitudes are $\sim\pm$0.1 (see also Figure 3).\label{fig4}}
\end{figure}
\clearpage

\begin{figure}
\epsscale{0.8}
\includegraphics[angle=90, width=17.5cm, height=13.0cm]{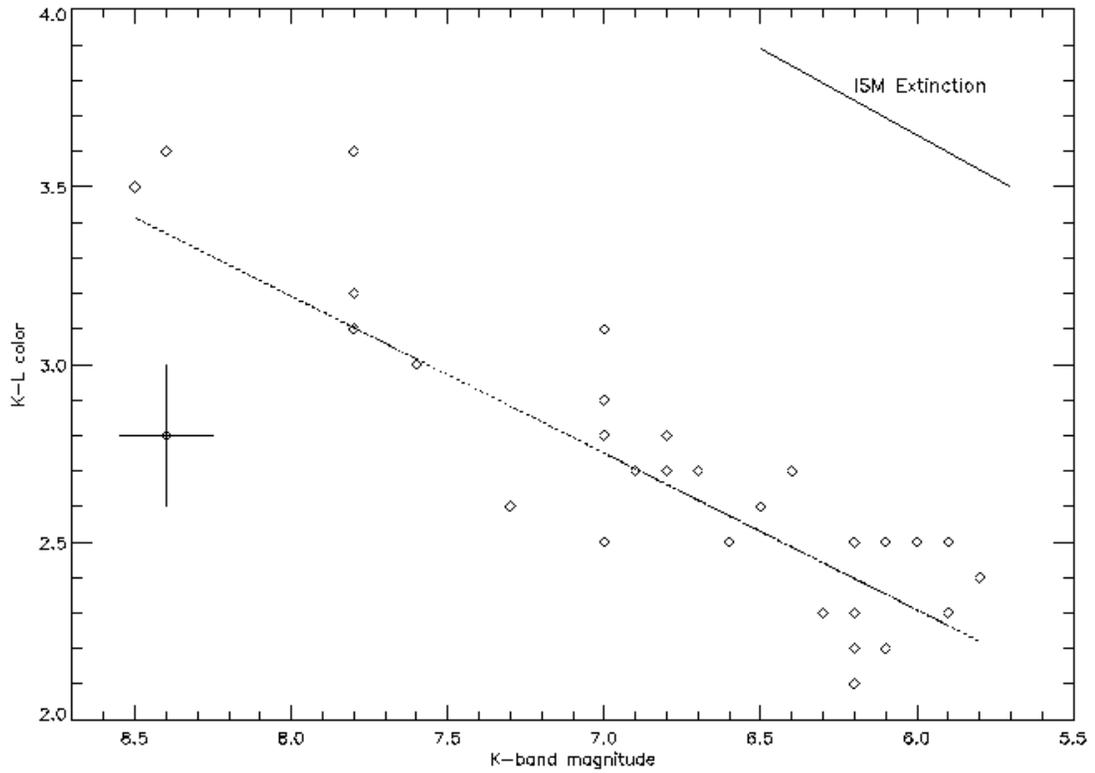}
\caption{The K-L$'$ color is plotted versus the simultaneous K-band magnitude for 38 observations of the T Tau S.  The point in the middle left shows an average measurement uncertainty for the data.  A correlation is seen between the color and brightness.  T Tau S appears redder when faint, as expected if changes in obscuration produce the observed variability. \label{fig5}}
\end{figure}
\clearpage

\begin{figure}
\epsscale{0.8}
\includegraphics[angle=90, width=17.5cm, height=13.0cm]{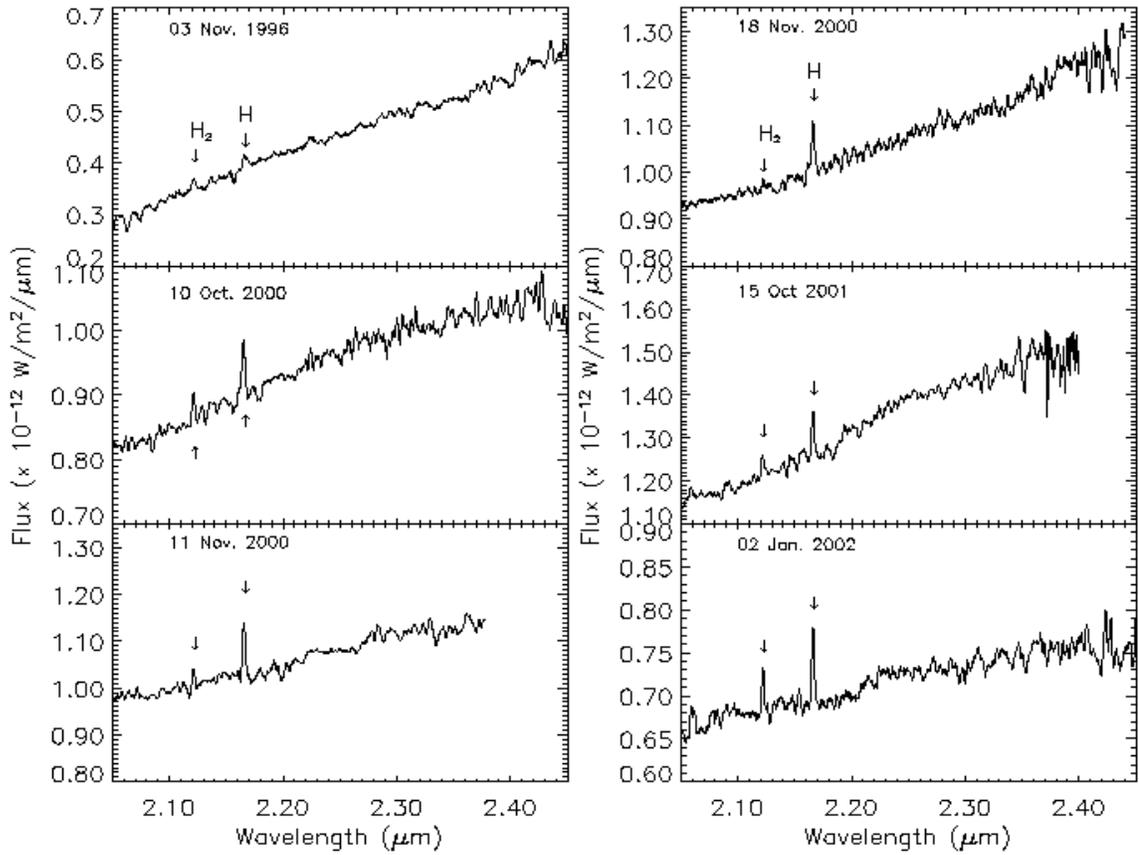}
\caption{The K-band (2.0-2.45 $\mu$m) spectra of T Tau S.  Emission of Hydrogen Br$\gamma$ and H$_2$ S=(1-0) lines are seen in the spectra.  We do not detect any photospheric absorption features characteristic of young solar type stars in the spectra of T Tau S.  \label{fig6}}
\end{figure}
\clearpage

\begin{figure}
\epsscale{1.0}
\includegraphics[angle=90, width=14.0cm, height=20.0cm]{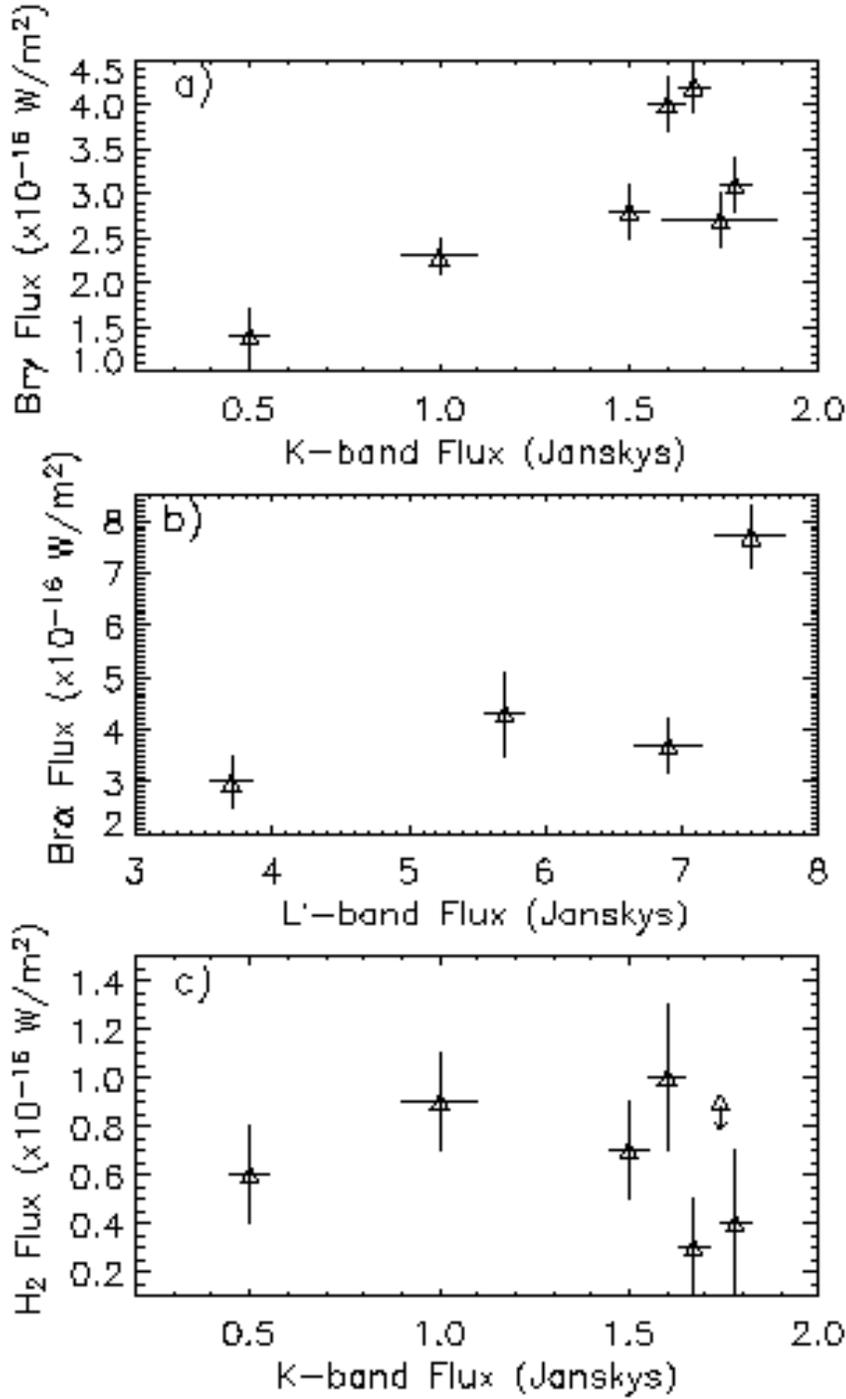}
\caption{Fluxes in the atomic and molecular hydrogen emission features of T Tau S plotted versus the simultaneous K and L$'$-band flux values.  The data show a tentative correlation between Br$\gamma$ emission and K-band flux over a factor of $\sim$3.5 variation in both values (a), at no time when T Tau S is bright is a weak Br$\gamma$ flux observed.  While there may also be a correlation between Br$\alpha$ and L$'$-band flux, fewer measurements are available to test this (b).  The H$_{2}$ emission appears to be uncorrelated with K-band flux (c). \label{fig7}}
\end{figure}
\clearpage

\begin{figure}
\epsscale{0.8}
\includegraphics[angle=90, width=17.5cm, height=13.0cm]{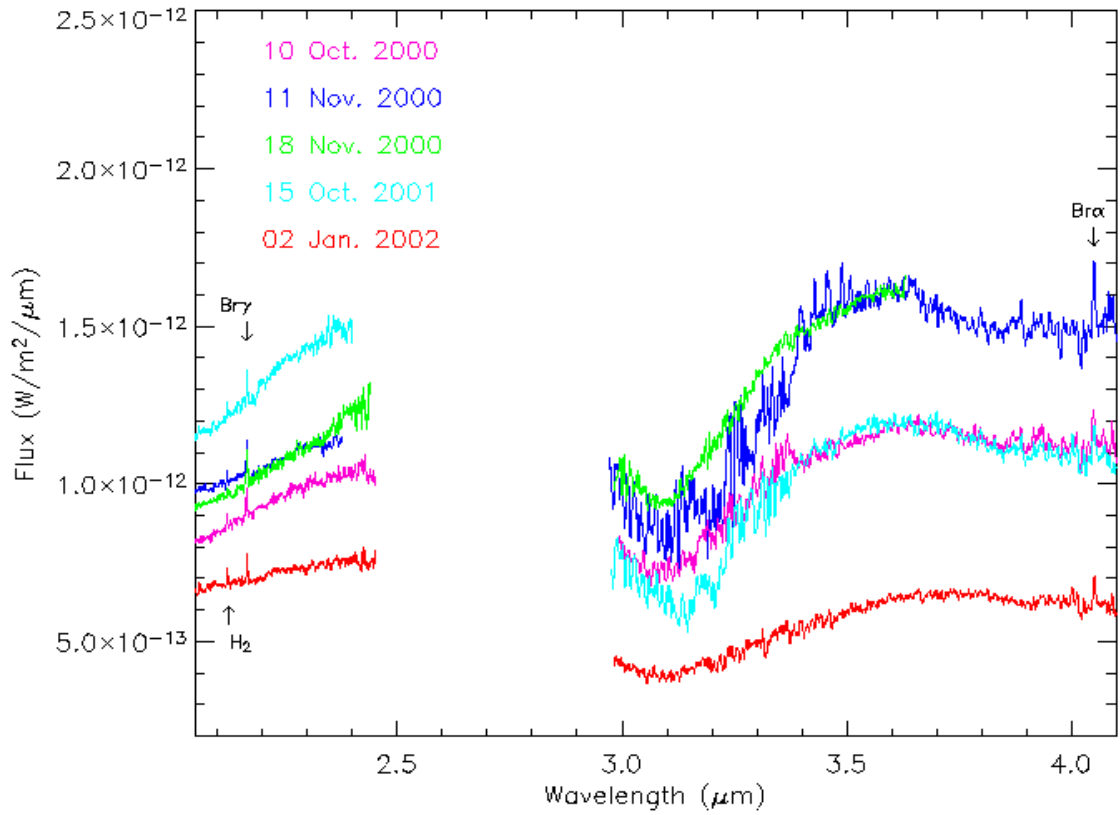}
\caption{The 2-4 $\mu$m spectra of T Tau S measured with SpeX on five occasions.  The broad absorption feature of water-ice, centered at $\sim$3.05$\mu$m, is obvious in all of the spectra.  Arrows indicate the H$_{2}$ {\it v}=1-0 S(1) (2.12 $\mu$m), hydrogen Br$\gamma$ (2.16 $\mu$m) and Br$\alpha$ (4.05$\mu$m) emission lines. \label{fig8}}
\end{figure}
\clearpage

\begin{figure}
\plotone{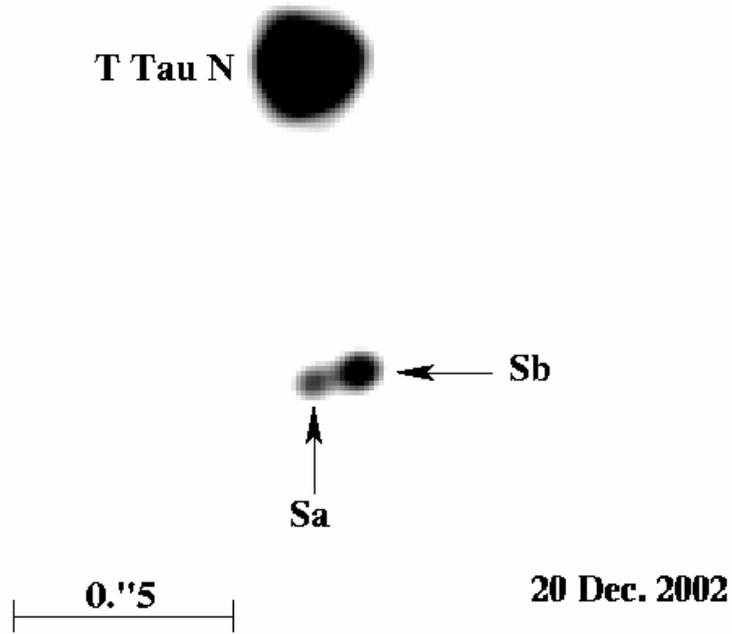}
\caption{A K-band image of the T Tau system from adaptive optics observations obtained at Gemini Observatory on 2002 Dec. 20.  T Tau Sa is modeled as the most luminous and massive component in the system, but is seen here as the faintest star in this K-band image. \label{fig9}}
\end{figure}
\clearpage

\begin{figure}
\epsscale{0.8}
\includegraphics[angle=90, width=17.5cm, height=13.0cm]{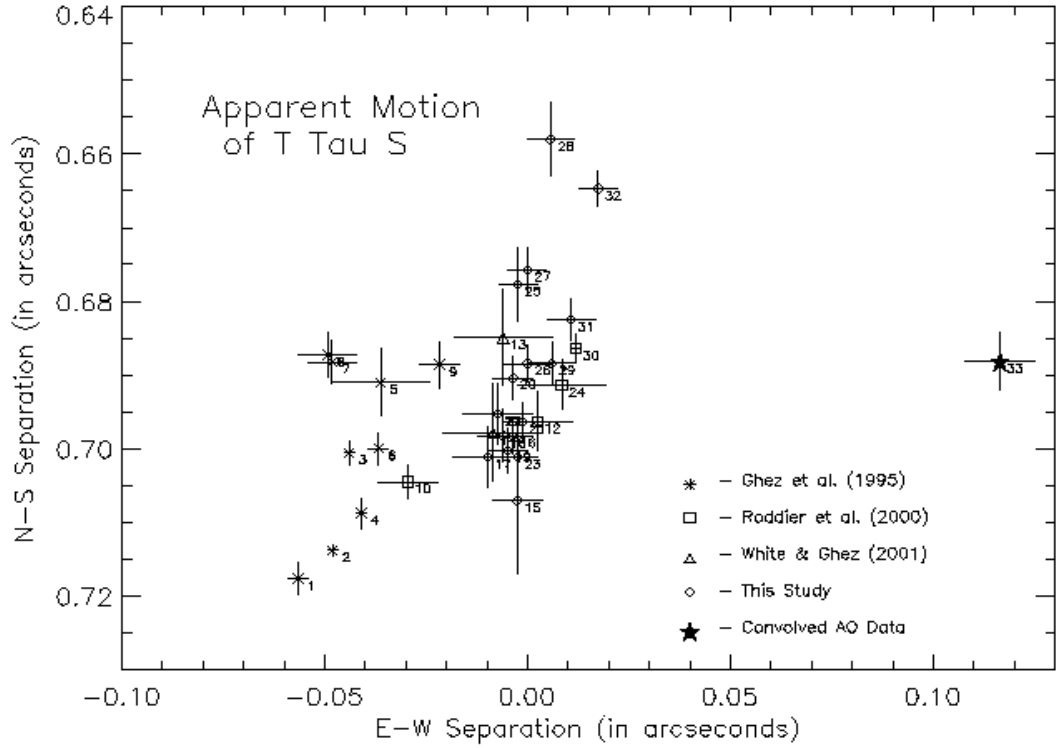}
\caption{The apparent motion of T Tau S with respect to T Tau N from K-band observations made 1989 to 2002.  The numbers overplotted to the lower right of each datapoint correspond to the observation number included in Table 6.  The position of T Tau N is off the top of the plot and corresponds to x=0.0 and y=0.0.  The star symbol represents AO data convolved to lower spatial resolution from images in which T Tau Sb is known to dominate the flux of T Tau S. \label{fig10}}
\end{figure}
\clearpage

\begin{figure}
\epsscale{0.8}
\includegraphics[angle=90, width=17.5cm, height=13.0cm]{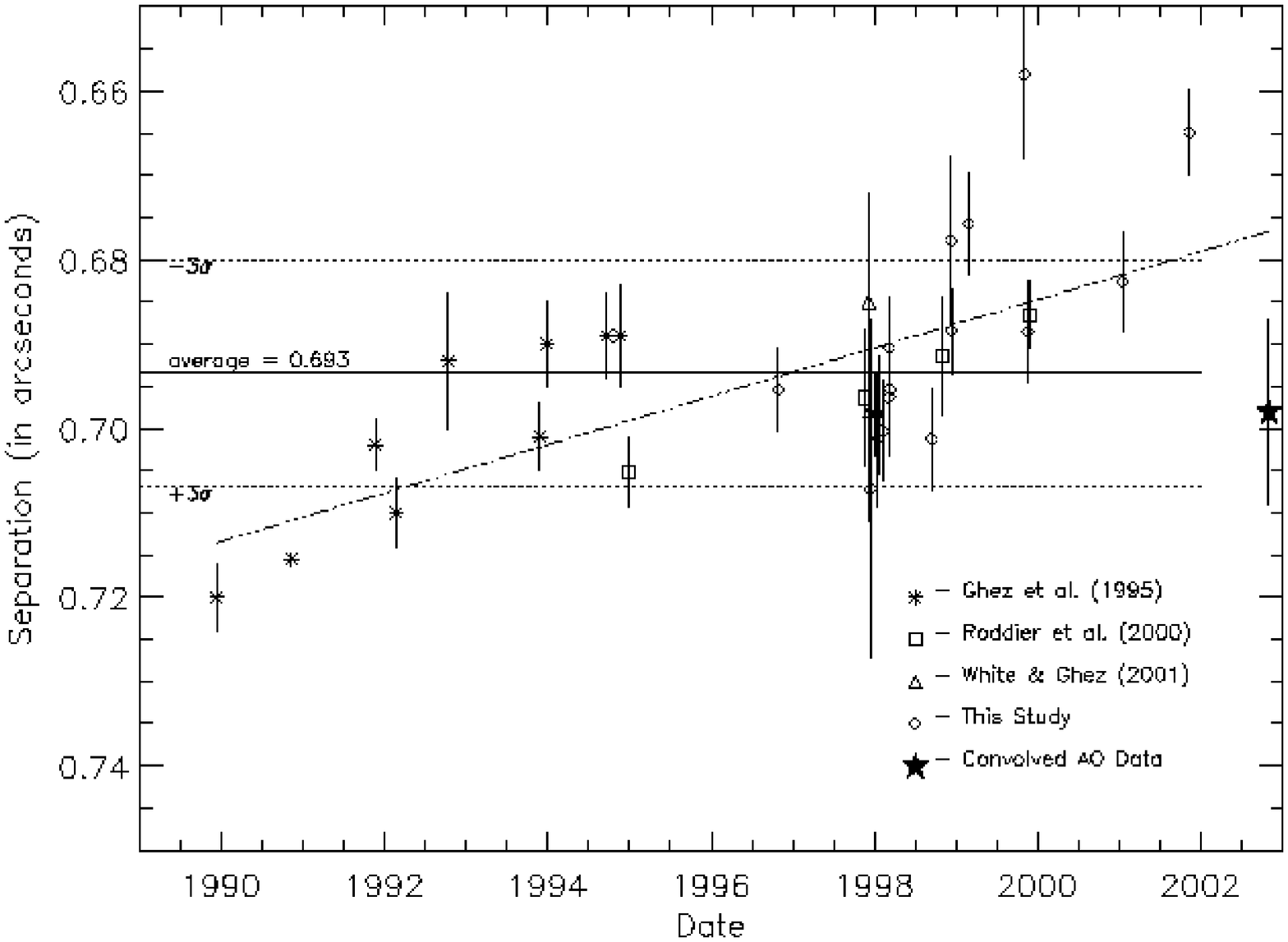}
\caption{The variation in the separation of T Tau S from T Tau N from K-band observations made from 1989 through 2002.  The dash-dot line is a linear fit to the apparent changes in separation versus time.  Over the 12 years of the observations, the separation has become smaller at a high level of significance.  The star symbol represents AO data convolved to lower spatial resolution from images in which T Tau Sb is known to dominate the flux of T Tau S. \label{fig11}}
\end{figure}
\clearpage

\begin{figure}
\epsscale{0.8}
\includegraphics[angle=90, width=17.5cm, height=13.0cm]{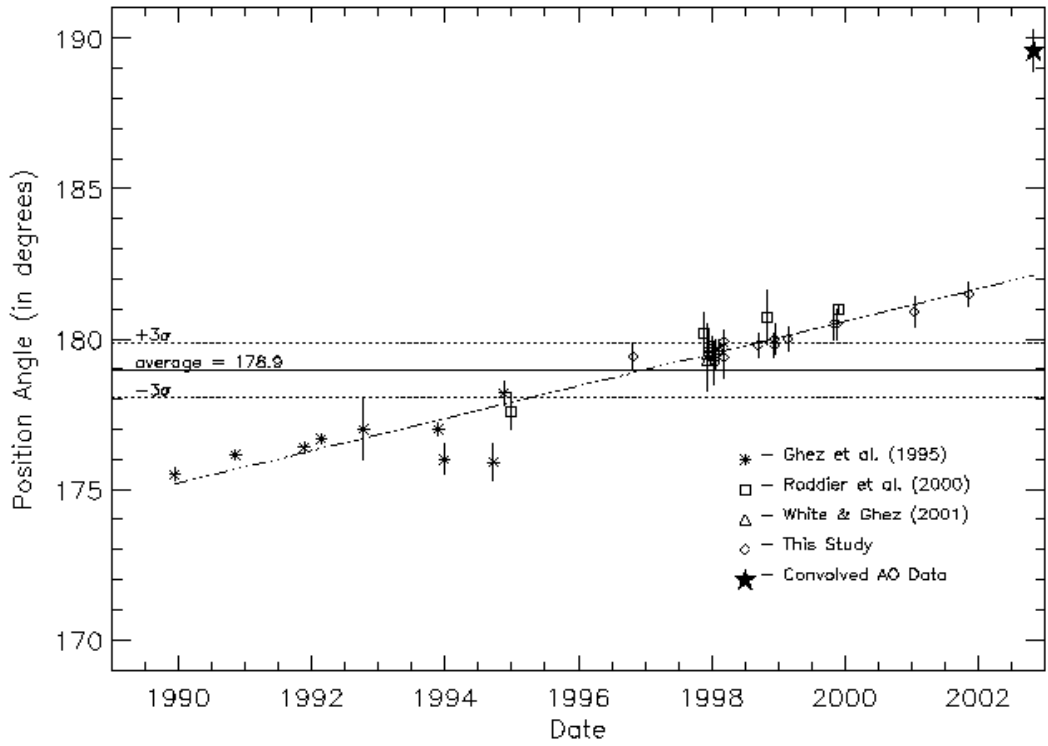}
\caption{The variation in the position angle of T Tau S with respect to T Tau N from K-band observations made from 1989 through 2002.  The dash-dot line is a linear fit to the apparent position angle versus time.  Over the 12 years of the observations, the position angle has increased significantly.  The star is from AO data convolved to lower spatial resolution from images in which T Tau Sb is known to dominate the flux of T Tau S. \label{fig12}}
\end{figure}
\clearpage

\begin{figure}
\plotone{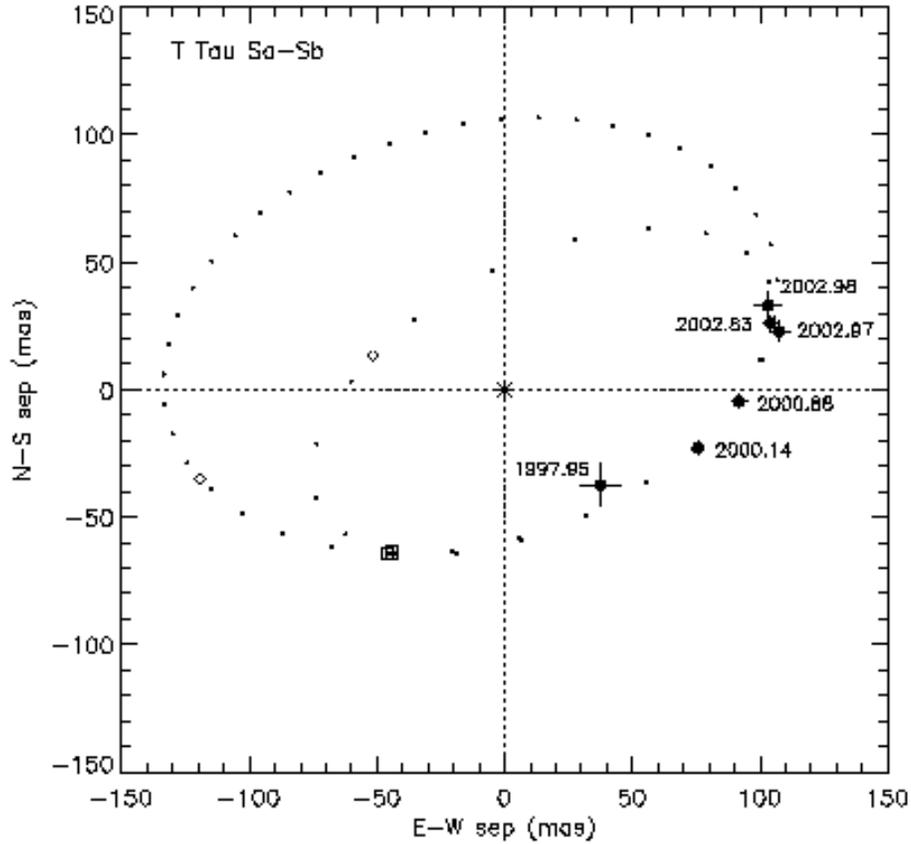}
\caption{The orbital motion of the T Tau Sa/Sb system from the high spatial resolution infrared observations centered T Tau Sa which is marked by an asterisk.  Overplotted in the figure are two possible orbital models, with periods of 20 and 40 years (plotted at yearly intervals).  The diamonds and squares correspond to the predicted positions of T Tau Sb for each orbit at the times, respectively, of the speckle observations of Ghez et al. (1991) in 1990.60 and the lunar occultation observation of Simon et al. (1996) made in 1994.96. \label{fig13}}
\end{figure}
\clearpage

\begin{figure}
\plotone{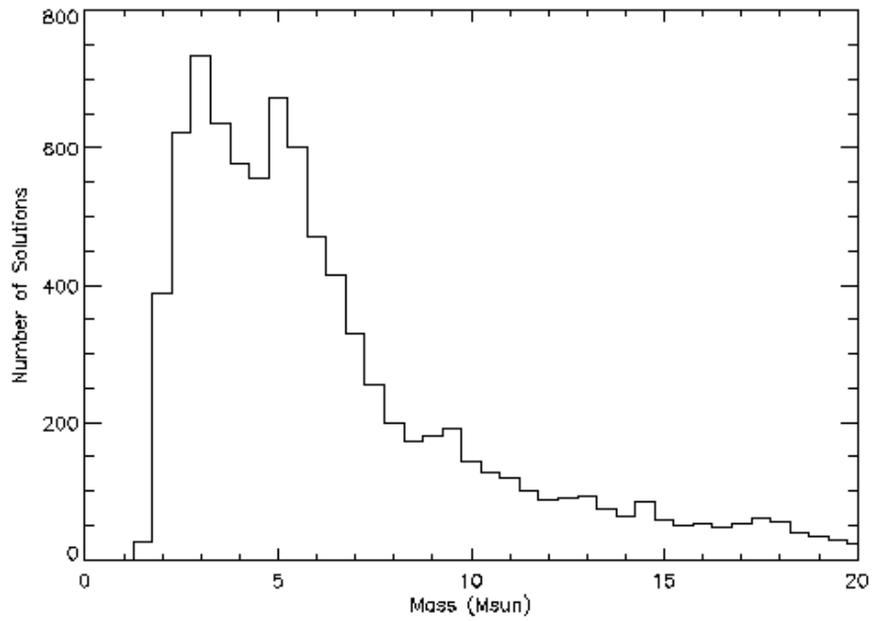}
\caption{A histogram of possible system masses for T Tau S derived from varying model parameters that fit orbits to the observations of the Sa-Sb system.  Our analysis finds that nearly 1/3 of the possible mass solutions are greater than 10 M$_{\odot}$.  However, these masses are not consistent with the luminosities derived for T Tau S.  \label{fig14}}
\end{figure}
\clearpage

\begin{figure}
\epsscale{0.8}
\includegraphics[angle=90, width=17.5cm, height=13.0cm]{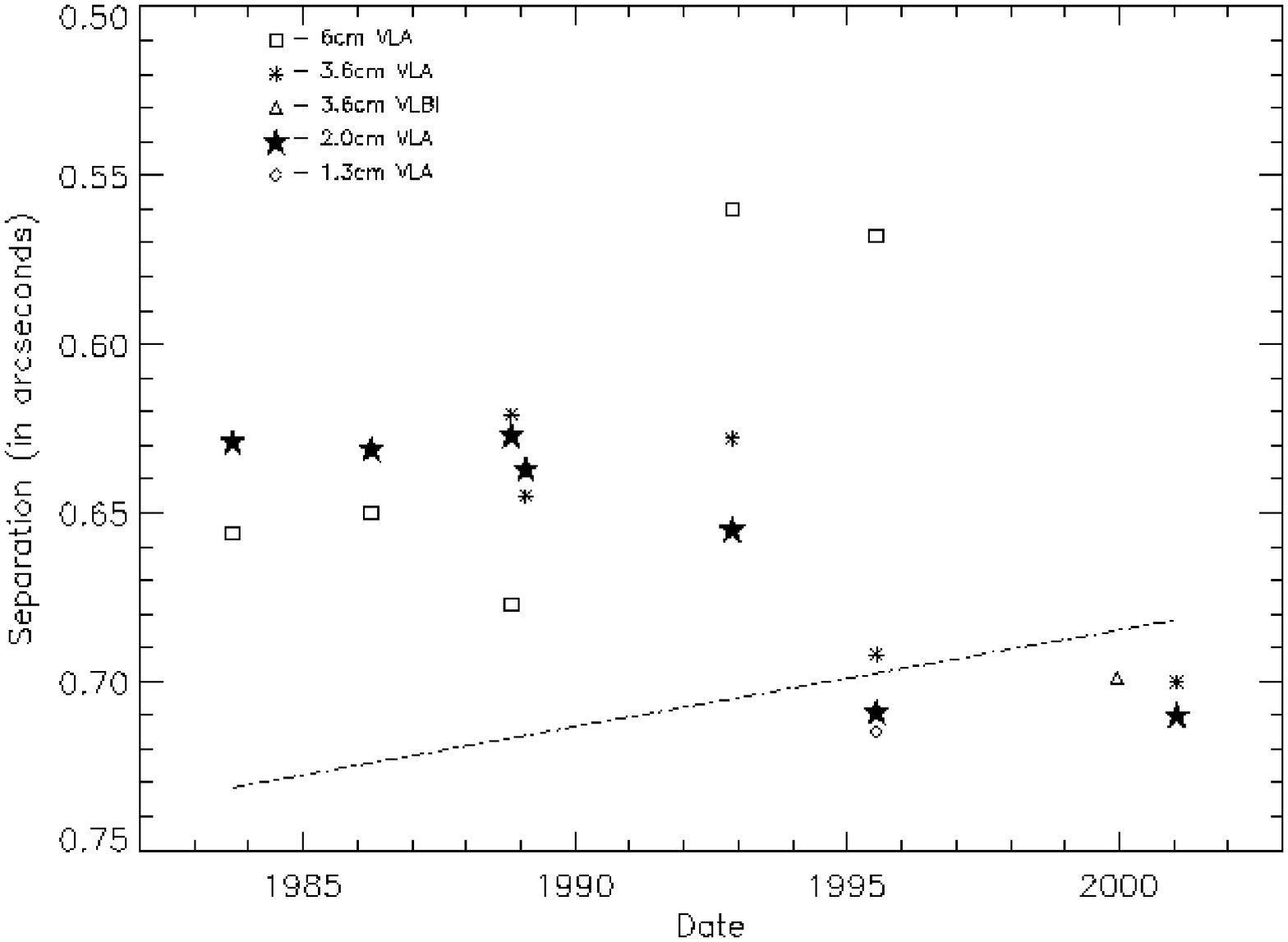}
\caption{The coded symbols plot the variation in the separation of the VLA radio source associated with T Tau S with respect to T Tau N.  The dashed line is the linear fit of the motion of T Tau S with respect to T Tau N from our infrared observations (from Figure 11). \label{fig15}}
\end{figure}
\clearpage

\begin{figure}
\epsscale{0.8}
\includegraphics[angle=90, width=17.5cm, height=13.0cm]{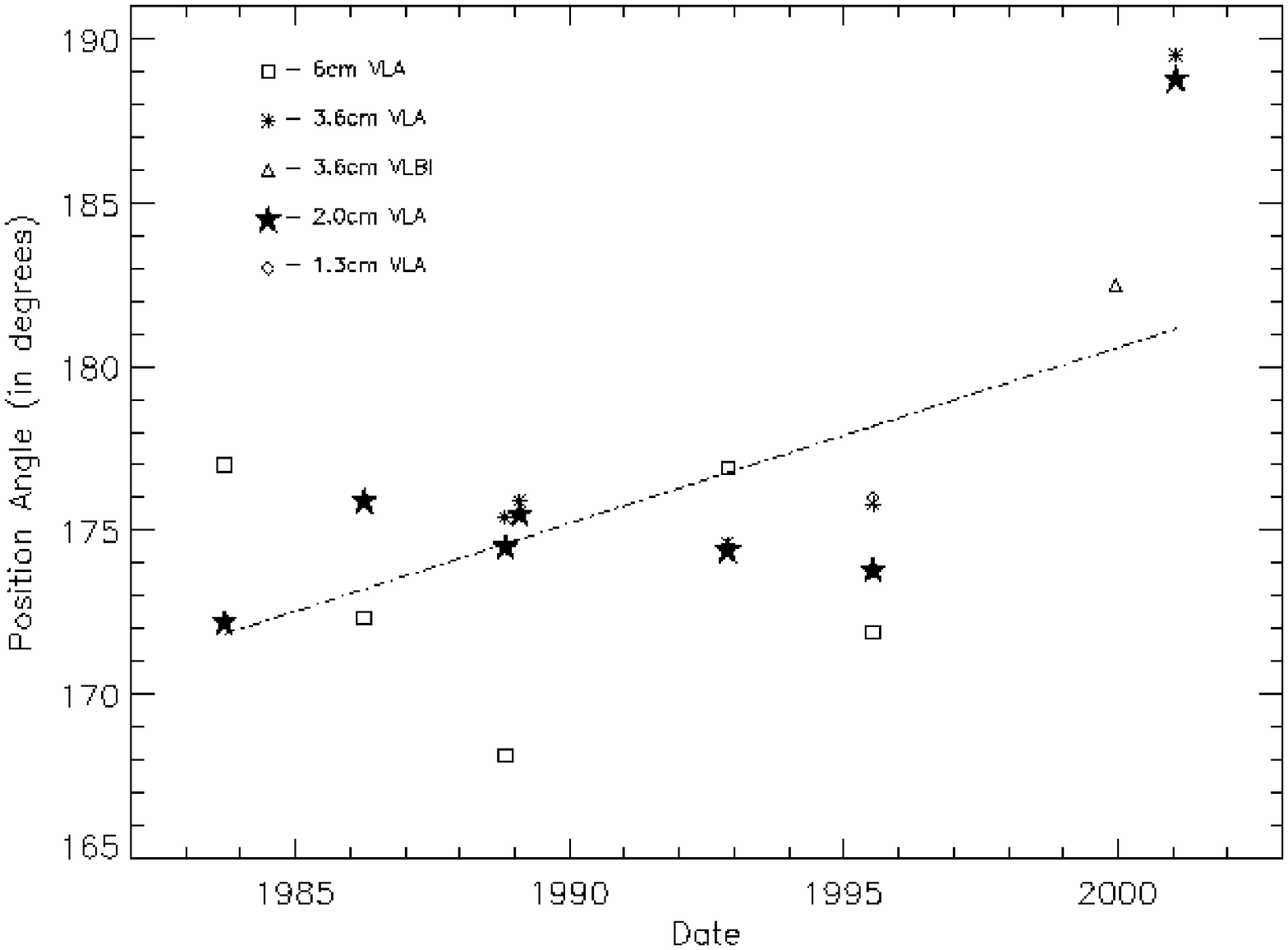}
\caption{The coded symbols plot the variation in the position angle of the VLA radio source associated with T Tau S with respect to T Tau N.  The dashed line is the linear fit of the motion of T Tau S with respect to T Tau N from our infrared observations (from Figure 12). \label{fig16}}
\end{figure}
\clearpage

\begin{figure}
\epsscale{0.7}
\includegraphics[angle=90,width=12.0cm,height=19.0cm]{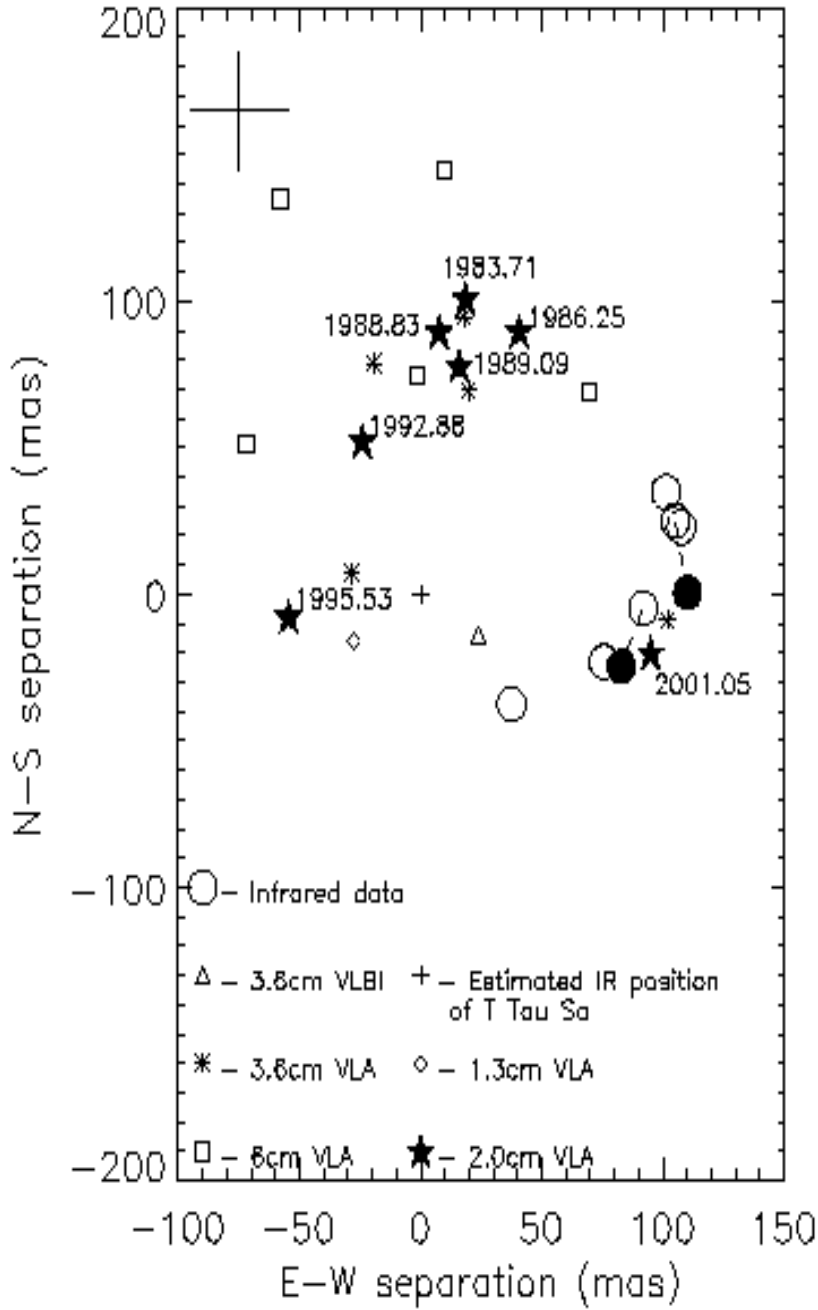}
\caption{The position of the radio source is plotted with respect to the average position of T Tau Sa derived from our study.   Included in the plot are the infrared positions of T Tau Sb from data reported in the literature and from observations described in this paper.  The filled circles are the estimated position of T Tau Sb derived from the infrared adaptive optics observations, but referencing the position of Sb to T Tau N and then using the same method applied to the radio data.  Overplotted in the upper left is the uncertainty of $\pm$20 mas that results from deriving positions using the average linear motion of T Tau Sa.  In the 2001 observations, the infrared position of T Tau Sb and the radio data lie within $\sim$15 mas of one another.  \label{fig17}}
\end{figure}
\clearpage


\end{document}